\documentclass[apj,a4]{emulateapj}

\pdfoutput=1

\usepackage{multirow}

\newcommand{\Mgb}{Mg\,\emph{b}}
\newcommand{\NaD}{Na\,\textsc{d}}
\newcommand{\Ha}{H$\alpha$}
\newcommand{\Hb}{H$\beta$}
\newcommand{\HII}{H\textsc{ii}}
\newcommand{\Nii}{[N\,\textsc{ii}]}
\newcommand{\Sii}{[S\,\textsc{ii}]}
\newcommand{\Siii}{[S\,\textsc{iii}]}
\newcommand{\Oi}{[O\,\textsc{i}]}
\newcommand{\Oii}{[O\,\textsc{ii}]}
\newcommand{\Oiii}{[O\,\textsc{iii}]}

\newcommand{\etal}{\emph{\,et\,al.}}
\newcommand{\kms}{\ km s$^{-1}$}

\shorttitle{Galactic Winds}
\shortauthors{Sharp \& Bland-Hawthorn}
\begin{document}

\title{3D Integral Field Observations of Ten Galactic Winds $-$
I. Extended phase ($\ga$10 Myr) of mass/energy injection before the wind blows}

\author{R.G. Sharp\altaffilmark{1}}
\altaffiltext{1}{Anglo-Australian Observatory, PO Box 296, Epping, NSW 1710, Australia}
\email{rgs@aao.gov.au}
\author{J. Bland-Hawthorn\altaffilmark{2,3}}
\altaffiltext{2}{Sydney Institute for Astronomy, School of Physics A28, University of Sydney, NSW 2006, Australia}
\altaffiltext{3}{Leverhulme Visiting Professor, Physics Department, University of Oxford, 1 Keble Rd, Oxford, OX1 3RH, UK}
\email{jbh@physics.usyd.edu.au}

\begin{abstract}
In recent years, we have come to recognize the widespread importance
of large-scale winds in the lifecycle of galaxies. The onset and
evolution of a galactic wind is a highly complex process which must be
understood if we are to understand how energy and metals are recycled
throughout the galaxy and beyond. Here we present 3D spectroscopic
observations of a sample of 10 nearby galaxies with the AAOmega-SPIRAL
integral field spectrograph on the 3.9m AAT, the largest survey of its
kind to date. The double-beam spectrograph provides spatial maps in a
range of spectral diagnostics: \Oiii 5007, \Hb, \Mgb, \NaD, \Oi 6300,
\Ha, \Nii 6583, \Sii 6717, 6731.  We demonstrate that these flows can
often separate into highly ordered structures through the use of
ionisation diagnostics and kinematics.  All of the objects in our
survey show extensive wind-driven filamentation along the minor axis,
in addition to large-scale disk rotation. Our sample can be divided
into either starburst galaxies or active galactic nuclei (AGN),
although some objects appear to be a combination of these.  The total
ionizing photon budget available to both classes of galaxies is
sufficient to ionise all of the wind-blown filamentation out to large
radius.  We find however that while AGN photoionisation always
dominates in the wind filaments, this is not the case in starburst
galaxies where shock ionisation dominates. This clearly indicates that
after the onset of star formation, there is a substantial delay ($\ga$
10 Myr) before a starburst wind develops. We show why this
behavior is expected by deriving ``ionisation'' and dynamical
timescales for both AGNs and starbursts. We establish a sequence of
events that lead to the onset of a galactic wind. The clear signature
provided by the ionisation timescale is arguably the strongest evidence
yet that the starburst phenomenon is an impulsive event. A well-defined
ionisation timescale is not expected in galaxies with a protracted
history of circumnuclear star formation.  Our 3D data provide
important templates for comparisons with high redshift galaxies.
\end{abstract}


\section{Introduction}

Recent observations reveal that outflows are important in the
lifecycle of galaxies. First, \citet{Tremonti_04} find evidence for
chemical enrichment trends throughout star-forming galaxies over three
orders of magnitude in stellar mass. Surprisingly, the enrichment
trends are observed to continue all the way down to $\log(L_{\rm
bol}/L_\odot) \approx 4$ \citep{Kirby_08} where $L_{\rm bol}$ is the
bolometric luminosity of the galaxy. Indeed, there is evidence that
galaxies that exceed the mass of the Milky Way manage to retain a
large fraction of their metals, in contrast to lower mass galaxies
where metal loss appears to be anticorrelated with baryonic
mass. Recent theoretical work has shown that a low-metal content in
dwarf galaxies is attributable to efficient metal-enriched outflows
\citep{Dalcanton_07}. Secondly, there is now strong evidence for
large-scale outflows in the Galaxy across the electromagnetic spectrum
\citep{Bland_03, Fox_05, Keeney_06}. Finally, the intergalactic medium
(IGM) at all redshifts shows signs of significant metal enrichment
consistent with the action of winds \citep{Cen_99, Madau_01, Ryan_06,
Dave_08}.  A high proportion of Lyman break galaxies at $z\sim 3-4$
show kinematic signatures of winds \citep{Erb_06} as do many nearby
dwarf starbursts \citep{Schwartz04} and ultraluminous infrared galaxies
\citep{Martin06,Schwartz06}.

N-body simulations of galaxy formation within cold dark matter (CDM)
cosmology fail to produce realistic galactic disks \citep{Navarro_94,
Steinmetz_95}.  Over-cooling in CDM+hydrodynamic simulations, due
primarily to a lack of resolution, results in too much gas arriving in
the centre of the galaxies.  But a useful aspect of these incomplete
models has been to highlight the possible role of feedback in shaping
galaxies. Vigorous feedback in the early phase of galaxy formation is
a partial solution to the angular momentum problem by preventing
baryons from losing too much of their specific angular momentum
through the interaction with dark matter \citep{Fall_02}. Indeed, some
authors have claimed more realistic disks retaining much more of their
angular momentum when one includes the action of AGN-driven jets
\citep{Robertson_06} or starburst driven winds \citep{Efstathiou_00}.

Our new survey is preempted by a vast literature across a broad
spectrum that can be traced back over two decades $-$ comprehensive
reviews are given by Heckman\etal\ (1989) and Veilleux\etal\
(2005). Much of the early optical work on galactic outflows was
limited to narrowband images and long-slit spectroscopy (e.g.\
McCarthy\etal\ (1987); \citet{Heckman90}; Phillips (1993); Lehnert \&
Heckman (1996)) although Fabry-Perot observations have been carried
out in a few emission lines (e.g.\ Bland \& Tully (1988); Cecil
(1990)). The quality of UV, x-ray, infrared and radio continuum
observations has greatly improved in recent years (e.g.\ Hoopes\etal\
2005; Strickland\etal\ 2004; \citet{Grimes09}).  Even without the
kinematic separation of compound emission structures that is possible
via spectroscopy, there are clear associations between these bands and
the optical maps (e.g.\ Veilleux\etal\ 1994; Cecil\etal\ 2002).
Historically, most winds had been confirmed kinematically from
evidence of line-splitting along the galaxy minor axis, but at least
one wind has been discovered through its ionisation signature
\citep{vr2002}.  In recent times measurement of blueshifted absorption
lines have proved powerful in this regard.

In this new study, we perform wide field integral field spectroscopy of
a sample of nearby galactic winds.  The advantages offered by a wide
wavelength base line and moderate spectral resolutions allow the
determination of physical parameters for wind outflow mechanics based
on ionisation and kinematic signatures e.g.\ rotating disk,
high-latitude warm ionizing medium, wind filaments, entrained gas.
One well-studied object in particular convinced us of the potential
for wide-field mapping with an integral field spectrograph. The
projected distribution of optical line emission in M82 is exceedingly
complex but with the aid of Fabry-Perot observations, Shopbell \&
Bland-Hawthorn (1998) were able to kinematically separate the front
side from the back side of the wind flow, and from the diffuse halo of
broadened line emission (see also the recent IFS studies of
\citet{Westmoquette09a, Westmoquette09b}).  When this is done, both
the ionisation and kinematics exhibit well ordered behavior.  The
lesson is clear: with complete observations, it is possible to
separate distinct dynamical components. Only then does it become
possible to determine physical parameters for the outflow mechanism.

In this first paper, we present the new observations and describe the
reduction procedures specific to integral field spectroscopy
(\S\,\ref{obsred}). We present emission-line flux and line ratio maps
for a wide range of diagnostics (e.g.\ ionisation and density).  For
each galaxy, we use the ionisation diagnostic diagrams (IDDs) to
delineate the different sources of ionisation as a function of
location and ionisation source (see \S\,\ref{the galaxy sample}) and
classify the sample targets according to their individual IDDs
(\S\,\ref{starbursts} - starbursts \& \S\,\ref{AGN} - AGN).  These
diagnostic diagrams reveal well-ordered behavior across each
galaxy. In \S\,\ref{AGN SF discussion} we consider the source of
ionisation from a theoretical perspective and use these considerations
to construct an evolutionary sequence for galactic winds in section
\S\,\ref{events}.  Our conclusions are presented \S\,\ref{discussion}.

\smallskip
Through this work, we assume a flat concordance cosmology such that
 H$_o$=71\,km\,s$^{-1}$\,Mpc$^{-1}$, $\Omega_M$=0.27 and
 $\Omega_{\Lambda}$=0.73.

\section{Observations and data reduction}
\label{obsred}

\subsection{Observing set up and procedure}

The AAOmega integral-field spectrograph at the AAT 3.9m is ideally
suited to our study (see Appendix).  The SPIRAL integral field unit (IFU) is a 32$\times$16 element
rectangular microlens array coupled via an optical fibre feed to the
dual beam AAOmega spectrograph \citep{sau2004,Sharp06}.  Mounted at
the Cassegrain auxilliary focus, the 0.7\arcsec\ IFU pixel scale is
well matched to the typical seeing at the AAT, and gives excellent
sensitivity to extended low surface brightness features.  The single
pointing field-of-view is 22.4\arcsec$\times$11.2\arcsec.  High accuracy, guide-probe
offset guiding allows simple and reliable mosaicing of large areas.
The default position angle of the IFU places the long axis in an east-west direction.
Observations of a number of targets were undertaken at alternate PAs
to accommodate the demands of other observing programs undertaken
alongside the observations reported here.

Observations were undertaken over a number of observing runs as
detailed in Table~\ref{obs runs}.  The dual beam AAOmega spectrograph
provides excellent coverage of the spectral features of interest,
while maintaining a high spectral resolution.  We observe with the
1500V and 1000R Volume Phase Holographic (VPH) gratings in the blue
and red arms respectively.  We use a 5700\AA\ dichroic beam splitter
and center the wavelength coverage and VPH blazes at 5000\AA\ and
6800\AA, resulting in 2.4 pixel mid-range resolution elements of
R$\sim$6300 and R$\sim$5600 (velocity resolutions of $\approx$ 35 and
$\approx$50 km s$^{-1}$ FWHM).  On accounting for the sample redshift
range, this targets the emission lines \Hb/\Oiii\ in the blue and
\Oi/\Ha/\Nii/\Sii\ in the red.
Unfortunately the important \Oii\ doublet at 3727\AA\ falls
outside of the observable spectral range with the chosen settings.
We consider this to be a crucial line for future studies of this kind.

Quartz-Halogen flatfield exposures and CuAr$+$FeAr arc lamp exposures
are taken at intervals throughout the observation to allow fibre tracing
on the CCDs (both arms of AAOmega use E2V 2k$\times$4k CCDs), flat
fielding and wavelength calibration.  Stationed in the west Coud\'e
room, AAOmega is kept thermally and gravitationally stable throughout each
night's observations in order to minimize the overhead on system
calibrations. The strong night sky oxygen line at 5577\AA\ falls outside the
observed range and so cannot be used to calibrate relative fibre
transmission as is the common practice in low resolution multi-object
spectroscopy.  Therefore twilight flatfield frames are observed in
order to account for relative fibre-to-fibre transmission variations.

An assortment of spectrophotometric standard
stars\footnote{Spectrophotometric standard stars were chosen from the
ESO spectrophotometric standard list available at
http://www.eso.org/sci/observing/tools/standards/spectra/} were
observed throughout each observing block, in order to prepare a flux
calibration solution.
The scatter in the calibration function derived from the standards
during the observations indicates a 10\% uncertainty in the absolute
flux calibration level.  The relative calibration with wavelength, a
quantity more relevant to line ratio determinations, is consistent at
the $\sim$2-3\% level, as determined from the comparison of
calibration observations over the course of the survey.

Each target was observed as part of a multi-pointing mosaic.
An overlap of 1 or 2 IFU elements (0.7\arcsec\ or 1.4\arcsec) was
allowed to enable relative intensity scaling and accurate mosaicing of
the final data sets.  Individual science exposures were typically
1200 sec.  Mosaic sizes vary between galaxies, but are typically of the
order of six to twelve IFU positions, yielding mosaics of the order
20\arcsec\ to 60\arcsec\ on a side.

Since our targets completely fill each IFU pointing, it is not
possible to generate a sky subtraction solution from fibres within
each IFU pointing.  To achieve accurate sky subtraction, a small number of
dedicated offset sky frames was taken each night.  Data was processed
in real time at the telescope using the \texttt{2dfdr}
software package.  This allowed monitoring of the observing conditions
such that the mosaic pattern could be repeated or adapted if required.
The final data analysis was carried out using a custom suite of
\textsc{idl} routines.

There are four isolated ``dead elements'' within the SPIRAL IFU.
During the reduction process, these are replaced with the average spectrum of 
the four adjacent IFU elements prior to mosaicing the data.  Individual frames
were then aligned and mosaiced using telescope offset information and
a suite of custom written software tools.
Frames are scaled, based on an iterative comparison of
overlap regions in the mosaic, in order to account for minor
variations in transparency, seeing and exposure time.

\begin{table}
\caption{\label{obs runs} A summary of AAT observations with the AAOmega$-$SPIRAL
spectrograph.}
\begin{center}
\begin{tabular}{ll}
\textbf{Observing Dates} & \textbf{Objects Observed}\\
\hline
\hline
May 2007 & NGC\,3628, NGC\,6810, NGC\,5128(CenA)\\
October 2007 & NGC\,1365, NGC\,1482, NGC\,1808\\
July 2008 & Circinus, NGC\,253, IC\,6063\\
August 2008 & NGC\,6810, NGC\,1705, NGC\,6240\\
\hline
\end{tabular}
\end{center}
\end{table}

\subsection{The sample}
We targetted a sample of ten galaxies with known galactic winds
(Veilleux\etal\ 2003; Heckman\etal\ 1990; Lehnert \& Heckman 1996).
The details of the sources are presented in Table~\ref{objects}.
Given that the aim of our study is to compare the properties of AGN
vs.\ starburst winds, we did not include a ``control'' sample of
non-wind galaxies. Integral field observations are extremely time
consuming and indeed IFU surveys of galaxies have rarely been
attempted before. Notable examples are the SAURON survey of early type
galaxies (e.g.\ de Zeeuw \etal\ 2002) and the SINS survey at higher
redshift \citep{Genzel08}.

We limit our survey to southern hemisphere sources with bolometric
luminosities in the range $L_{\rm bol} = 1-10\times 10^{10}$ $L_\odot$
to ensure the impact of photoionisation by the central source (whether
starburst or AGN) is detectable out to large galactocentric
distances. (The physical motivation for this statement is given in \S
4.1.) With a view to going after a lower luminosity sample, we
attempted a preliminary study of the dwarf galaxy NGC 1705
\citep{Annibali03,Meurer92,Hunter93}.  Our initial observations (not
presented here) served to emphasize that low mass galaxies typically
have lower metallicities which render many of the ionisation
diagnostics intrinsically weak. A survey sample of a substantial
number of low-luminosity wind sources ($\ga$10) will require a
suitable IFU spectrograph on an 8m class telescope.

\begin{figure*}
\begin{center}

\includegraphics[width=50mm]{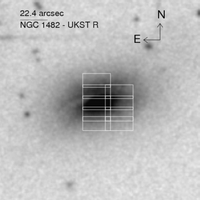}
\includegraphics[width=50mm]{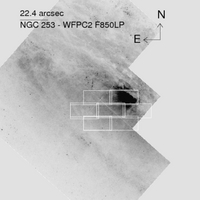}
\includegraphics[width=50mm]{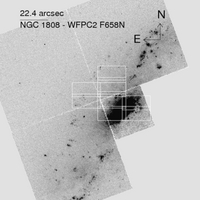}\\
\includegraphics[width=50mm]{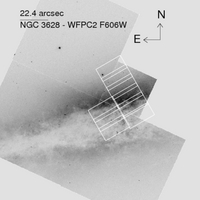}
\includegraphics[width=50mm]{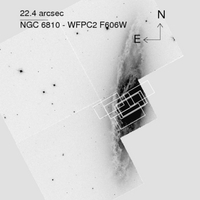}\\
\includegraphics[width=50mm]{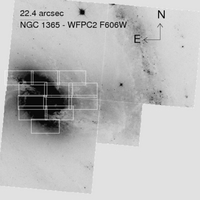}
\includegraphics[width=50mm]{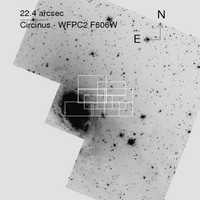}
\includegraphics[width=50mm]{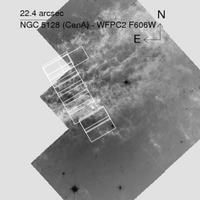}
\includegraphics[width=50mm]{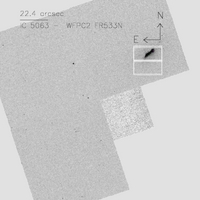}
\includegraphics[width=50mm]{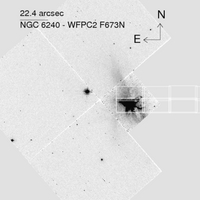}
\end{center}
  \caption{\label{finding charts} SPIRAL IFU footprints.  Images are
  taken from \emph{HST}/WFPC2 associations where available and from
  the DSS for NGC~1482.  The 22.4\arcsec\ long axis of the SPIRAL IFU is
  indicated for scale.  Note the unusual PA for observations of
  NGC~6810, NGC~3628 \& NGC~5128 due to the detailed requirements of
  the NGC~5128 observing program during the May 2007 run.}
\end{figure*}

For each of the ten wind galaxies, the SPIRAL IFU was used in a simple
tiling pattern to cover most of the filamentary emission above a
surface brightness of $1\times 10^{-17}$ erg cm$^{-2}$ s$^{-1}$
arcsec$^{-2}$ (cgs).  The SPIRAL mosaic patterns (or footprints) are
overlayed on \emph{HST}/WFPC2 imaging data in Fig.~\ref{finding
charts}.  WFPC2 images have been used for all sources with the
exception of NGC~1482 for which \emph{UKST} data from the Digitized
Sky Survey has been used. The \emph{HST}/WFPC2 data were obtained from
the Multimission Archive at the Space Telescope Science Institute
(MAST). Fig.~\ref{finding charts} emphasizes the need for much wider
field IFUs for the study of nearby galaxies. While these are under
development in Europe and Australia, we stress the simultaneous
requirements of spectral coverage and resolution (see Appendix) which
are often sacrificed in order to achieve a larger spatial format. In
this respect, we consider the AAOmega-SPIRAL spectrograph to be ideal
for the proposed science, even with the limited field of view.

\begin{table*}
\caption{\label{objects}  The target sample of southern or equatorial galactic wind sources.}
\begin{center}
\begin{tabular}{llllllllll}
\textbf{Object} &\textbf{RA/Dec (J2000)} & \textbf{Redshift} &\textbf{Hubble type}$^1$ & 
\textbf{Spec class}$^2$ &
\textbf{i} & \textbf{M$_{\rm{B}}$} & \textbf{L$_{\rm{Bol}}$$^{\rm{3}}$} &\textbf{R/Mpc} & 
\textbf{kpc/arcsec}\\
\hline
NGC~253  & 00 47.6 $-$25 18  & 0.00081 & Sc(X)   & \HII & 86 & $-$20.02 & 
2.8 &  2     & 0.017\\
NGC~1365 & 03 33.7 $-$36 08 & 0.00546 & Sb(B)   & \HII, Sy2   & 63 & $-$21.26 & 
9.3 & 16.9   & 0.100\\
NGC~1482 & 03 54.7 $-$20 30 & 0.00639 & Sa(P)   & \HII         & 58 & $-$18.89 & 1.1 
& 19.6   & 0.121\\
NGC~1808 & 05 07.7 $-$37 31  & 0.00332 & SO/a(X) & \HII, Sy2?          & 50 & $-$19.52 & 
2.2 & 10.8   & 0.059\\
NGC~3628 & 11 20.3 $+$13 37 & 0.00281 & Sb(P)   & \HII, LINER   & 87 & $-$19.96 & 
2.8 &  7.7   & 0.041\\
NGC~5128 & 13 25.3 $-$43 01  & 0.00183 & SO, Lenticular & Sy2    & 43 & $-$20.97 & 
8.5 &  4.9   & 0.053\\
Circinus & 14 13.2 $-$65 20       & 0.00145 & Sb(A)   & Sy2          & 65 & $-$21.23 & 9.0 &  
4.2   & 0.014\\
NGC~6240 & 16 53.0 $+$02 24 & 0.02445 & I0      & LINER, Sy2   & ---& $-$21.30 & 
8.2 & ---    & 0.5\\
NGC~6810 & 19 43.6 $-$58 40  & 0.00678 & Sa(A)   & \HII, Sy2?  & 82 & $-$20.61 & 5.5 
& 25.3   & 0.135\\
IC\,5063 & 20 52.0 $-$57 04      & 0.01135 & Sa      & Sy2   & ---& $-$20.34 & 4.2 & ---    
& 0.22\\
\hline
\hline
\end{tabular}
\end{center}
Note 1: Vizier database at http://vizier.u-strasbg.fr/\\
Note 2: NED database at http://nedwww.ipac.caltech.edu/; a question mark indicates that
the Seyfert classification has been questioned in later studies.\\
Note 3: Bolometric luminosities, L$_{\rm{Bol}}$, in units of 10$^{10}$ L$_\odot$,
are computed from the modeling of \citet{Buzzoni05}, with uncertainties
of typically 10\% for different assumed ages of the stellar population over the full range 1-14 Gyr.
\end{table*}

\begin{table}
\caption{\label{class} Classification of the ionisation mechanism
for the galactic wind emission.}
\begin{center}
\begin{tabular}{llll}
Object & \Oiii\ \& \Nii & \Oiii\ \& \Sii & \Oiii\ \& \Oi\\
\hline
\multicolumn{2}{l}{\textbf{Starbursts}}\\
NGC~253   & 10:90 & 10:90 & 10:90\\
NGC~1482 & 10:90 & 10:90 & 10:90\\
NGC~1808 & 20:80 & 0:0  &  0:0\\
NGC~3628 & 0:0$^1$ & ---  & ---\\
NGC~6810 & 0:0$^1$ & 0:0 & ---\\
\hline
\multicolumn{2}{l}{\textbf{AGNs}}\\
NGC~1365 &  90:10 & 90:10  & 100:0\\
NGC~5128 &  50:50 &  0:100 &  60:40\\
Circinus & 100:0  & 60:40  & 100:0\\
NGC~6240 & ---    & ---    & ---\\
IC~5063 & 100:0  & 90:10  & 100:0\\
\hline
\hline
\end{tabular}
\end{center}
\smallskip
The primary ionisation source in the galactic wind-driven filaments is
determined from the IDDs in Figs.~\ref{ngc1482 VO},\ref{ngc253
VO},\ref{ngc1808 VO},\ref{ngc3628 ratios},\ref{ngc6810
VO},\ref{ngc1365 VO},\ref{circinus VO},\ref{CenA VO} \& \ref{ic5063
VO}. For each of the IDDs, we show the percentage of points
(AGN:Shock) outside of the maximal star formation boundary
\citep{kew2001} consistent with the fiducial models derived from
NGC~1365 and NGC~1482.  Reliable fractions are hard to derive
so the numbers shown are indicative. A ratio of 0:0 indicates that all emission is
contained within the star formation boundary. Blank entries indicate
that the IDD could not be constructed in the given line ratio.\\ Note
1: Figs.~\ref{ngc3628 ratios} and \ref{ngc6810 ratios} clearly
delineate enhanced \Nii/\Ha\ co-spatial with line profile splitting
in the outflowing gas, consistent with
the shock signatures seen in NGC~1482 (see \S 3.1).
\end{table}

\begin{table*}
\caption{\label{gas extent} Spatial extent and brightness of the emission-line gas. }
\begin{center}
\begin{tabular}{lcccll}
\textbf{Object} & \textbf{L$_{\rm{Bol}}$} & \textbf{Extent} & \textbf{Deprojected} & \textbf{$\mu$(H$\alpha$) @ ~1$\;$kpc} & \textbf{$L$(H$\alpha$) @~1$\;$kpc} \\
& \textbf{(10$^{10}$\,L$_{\odot}$)} & \textbf{(arcsec)} & \textbf{(kpc)} & 
\textbf{(ergs\,s$^{-1}$\,cm$^{-2}$\,arcsec$^{-2}$)} &
\textbf{(ergs\,s$^{-1}$\,arcsec$^{-2}$)}\\
\hline
NGC~253   & 2.8 & 30.5  & 0.5  & 3.8e-16 $\pm$ 4.6e-16  & 5.4e+35 $\pm$ 6.5e+35 \\
NGC~1365  & 9.3 & 16.4  & 1.8  & 2.6e-16 $\pm$ 1.2e-16  & 1.4e+37 $\pm$ 6.5e+36 \\
NGC~1482  & 1.1 & 15.5  & 2.2  & 8.9e-16 $\pm$ 4.4e-16  & 6.6e+37 $\pm$ 3.3e+37 \\
NGC~1808  & 2.2 & 22.5  & 1.7  & 1.6e-16 $\pm$ 5.1e-17  & 3.7e+36 $\pm$ 1.2e+36 \\
NGC~3628  & 2.8 & 28.0  & 0.4  & 5.2e-17 $\pm$ 2.2e-17  & 1.7e+36 $\pm$ 6.9e+35 \\
NGC~5128  & 8.5 & 36.0  & 2.8  & 4.7e-16 $\pm$ 2.2e-16  & 6.8e+36 $\pm$ 3.2e+36 \\
Circinus  & 9.0 & 24.6  & 0.4  & 1.9e-16 $\pm$ 8.5e-17  & 1.5e+36 $\pm$ 6.6e+35 \\
NGC~6810  & 5.5 & 12.0  & 16.4 & 5.5e-16 $\pm$ 2.4e-16  & 4.7e+37 $\pm$ 2.1e+37 \\
IC\,5063  & 4.2 & 6.0   & 1.4  & 1.3e-15 $\pm$ 5.3e-16  & 3.1e+38 $\pm$ 1.3e+38 \\
\hline
\hline
\end{tabular}
\end{center}
The spatial extent of emission-line gas is estimated from measurements
of the \Ha\ and \Nii\ emission along the minor axis of each
galaxy. The point at which line fitting becomes unreliable is
determined for each mosaic.  For the default observing mode, this
corresponds to a line surface brightness of $2\times 10^{-17}$ cgs.
The physical extent within the mosaics is corrected for the galactic
inclination given in Table \ref{objects}.  In the final two columns,
we give the mean H$\alpha$ surface brightness and the mean H$\alpha$
luminosity at 1 kpc (with $1\sigma$ uncertainties), or at the physical
limit of the gas if this is less than 1 kpc. These values are not
corrected for internal dust extinction.
\end{table*}

\subsection{Emission line maps}
Emission line maps are constructed via profile fitting to each
individual spectrum from the data cubes.  Lines are approximated with
single Gaussian profiles in the work that follows, and each line is
fitted independently, rather than simultaneously, which is possible at
the SPIRAL resolution in most instances.  This simplification
overlooks effects such as line splitting observed in emission from the
front and back surfaces of an outflow cone and asymmetric profiles
(blue/red shoulders to emission lines) from marginally resolved
kinematic components.  Ultimately a full deconvolution of each data
cube, informed by kinematic models for each source, will be required
(e.g.\ \citet{Westmoquette09a}).

Currently no account is being taken for the underlying stellar
absorption although \Hb\ absorption is evident close to the disk in
some cases.  The limited wavelength coverage, which excludes Balmer
lines higher than \Hb, hampers a proper treatment of the underlying
stellar continuum.  The dominant effect of neglecting stellar
absorption will be to underestimate the flux in the \Hb\ emission
line, suggesting that \Oiii/\Hb\ ratios may be moderately
overestimated in regions with a significant contribution from stellar
continuum. In our present study, Balmer absorption has negligible
impact since we are concerned with filaments that extend beyond the
projected stellar disk.

\subsection{Ionisation diagnostic diagrams}
\label{IDD plots text}
For each object, we construct ``ionisation diagnostic diagrams''
(IDDs) as prescribed by \citet{bpt} and \citet{vo} for different
families of optical emission lines.  An example, constructed from the
observations of NGC~1482 (Figs.~\ref{ngc1482 lines} \& \ref{ngc1482
ratios}), is presented in Fig.~\ref{ngc1482 VO}.  We have included the
\emph{extreme starburst} model of \citet{kew2001} as a common fiducial
indicator. The IDDs for the galaxy sample are presented in
Figs.~\ref{ngc1482 VO}, \ref{ngc253 VO}, \ref{ngc1808 VO}, \ref{ngc6810
VO}, \ref{ngc1365 VO}, \ref{circinus VO}, \ref{CenA VO} \& \ref{ic5063
VO}.

We have added two further reference lines to each diagnostic plot,
which are derived from two objects in our sample.  The two diagonal lines
across the IDDs indicate the different tracks
traced by the bulk of the data points in NGC~1365 and NGC~1482. These
two galaxies are the clearest examples within our sample of AGN
photoionisation \citep{Veron80} and shock ionisation \citep{vr2002}
respectively.  As such, these tracks constitute our own internal
calibration of two very different sources of gaseous ionisation and
allow for a very precise differential analysis for our galaxy
sample. This clear distinction arising from the IDDs for each galaxy
allows us to cleanly separate most of the galaxies into one or other
category in terms of the dominant source of ionisation in the outer
filaments.

We emphasize that shock diagnostics can overlap a region of the IDDs
occupied by so-called LINER objects, some of which may be excited
by dilute photoionizing sources rather than by shocks. The origin of
LINER ionization is still somewhat controversial. For all of our
objects, the strength of the [NII], [SII] and [OIII] lines, and sometimes [OI] lines, 
argues against ``LINER'' ionization so we do not consider it here.

\section{The galaxy sample}
\label{the galaxy sample}

We provide an overview of each galaxy based on an extensive literature
review with particular emphasis on recent work.  Our initial
classification below in terms of starburst or AGN is taken from Table
2. We start with the canonical objects NGC~1482 and NGC~1365 in each
category respectively.

\subsection{Starbursts}
\label{starbursts}

\smallskip
\subsubsection*{NGC~1482}
\citet{Hameed99} first reported ``filaments
and chimneys of ionised gas'' in NGC~1482.  More recently, tunable
filter observations of NGC~1482 led \citet{vr2002} to consider it
the prime example of a ``shock excited, limb-brightened conic wind
structure''.  No compelling evidence for an AGN has yet been reported
in the literature at any wavelength and thus NGC~1482 is classified as a
typical \HII\ galaxy \citep{kew2001}.

\begin{figure*}
\begin{center}
  \includegraphics[width=50mm]{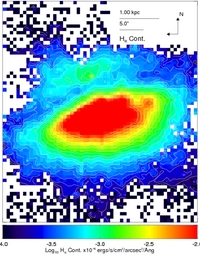}
  \includegraphics[width=50mm]{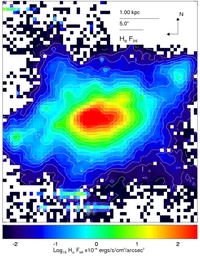}
  \includegraphics[width=50mm]{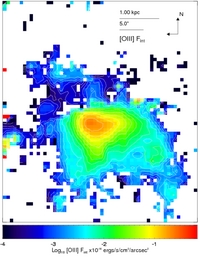}\\
  \includegraphics[width=50mm]{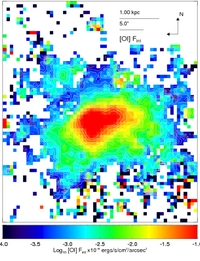}
  \includegraphics[width=50mm]{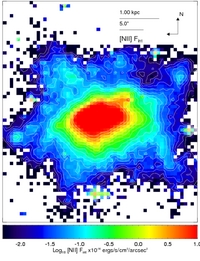}
  \includegraphics[width=50mm]{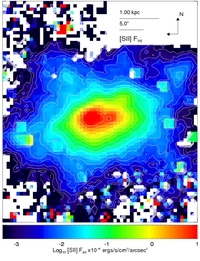}
\end{center}
  \caption{\label{ngc1482 lines} NGC~1482: Emission line integrated
  flux maps are genearted by single gaussian fitting to individual
  spectra. The maps shown (left to right, top to bottom) are the
  continuum at line centre for \Ha\ followed by integrated intensity
  maps (F$_{\rm{int}}$) for \Ha, \Oiii$\lambda$5007, \Oi$\lambda$6300,
  \Nii$\lambda$6583 \& \Sii$\lambda$6716. Pixels for which no valid
  fit was obtained are left blank.}
\end{figure*}

\begin{figure*}
\begin{center}
\includegraphics[width=50mm]{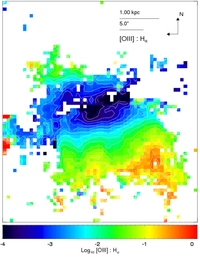}
\includegraphics[width=50mm]{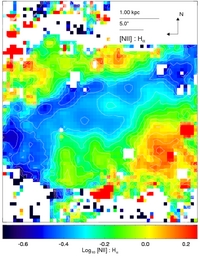}
\includegraphics[width=50mm]{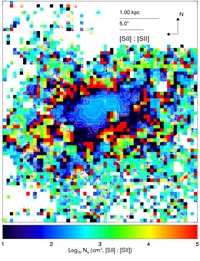}\\
\includegraphics[width=50mm]{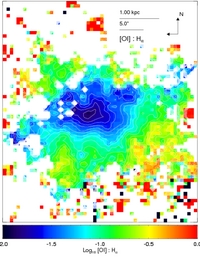}
\includegraphics[width=50mm]{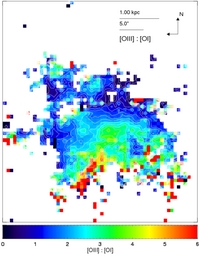}
\end{center}
  \caption{\label{ngc1482 ratios} NGC~1482: Line ratio maps are
  constructed from the emission line fits of figure \ref{ngc1482 lines}.
  The \Sii\ ratio is presented as an electron density, N$_e$,
  following \citet{Osterbrock}.  Pixels without valid measurements in
  both lines are left blank.}
\end{figure*}

\begin{figure*}
\begin{center}
\includegraphics[width=50mm]{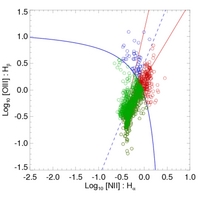}
\includegraphics[width=50mm]{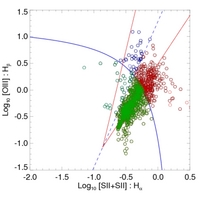}
\includegraphics[width=50mm]{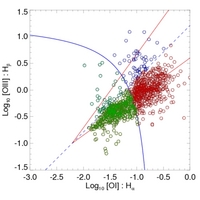}\\
\includegraphics[width=50mm]{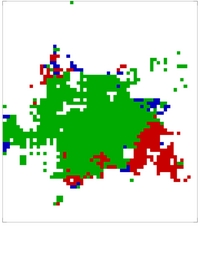}
\includegraphics[width=50mm]{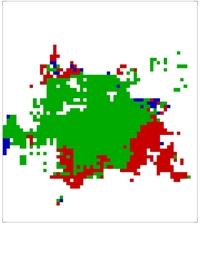}
\includegraphics[width=50mm]{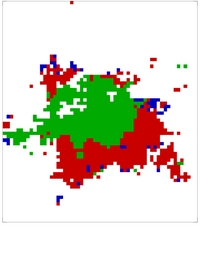}
\end{center}
\vspace{-1cm}
  \caption{\label{ngc1482 VO} NGC~1482: The classic Ionisation
  Diagnostic Diagrams (IDDs, see section \ref{IDD plots text}) are
  constructed from the integrated line flux measurements. Only IFU
  mosaic elements with valid line fits for all four (five for \Sii)
  lines are shown in each case. For each set of line ratios in turn,
  classification zones are defined and used to reconstruct the spatial
  distribution of the corresponding gas in the lower figures.  The
  curved solid line marks the \emph{extreme starburst} limit of
  \citet{kew2001}.  Two fiducial indicators are drawn (straight solid
  lines).  The right hand line traces the locus of points for the
  shock excited emission of NGC~1482 (this galaxy).  The left
  indicator traces the locus of line ratios for the AGN excited
  emission of NGC~1365 (Fig.~\ref{ngc1365 VO}).  The dashed line marks
  the bisector between these two fiducial traces.}
\end{figure*}

Our SPIRAL IFU observations of NGC~1482 are presented in
Figs.~\ref{ngc1482 lines}$-$\ref{ngc1482 VO}.  The conic outflow
structure seen in the emission-line gas \citep{vr2002} and diffuse
x-ray emission \citep{Strickland2004a} is evident in the SPIRAL
integrated line flux maps (Fig.~\ref{ngc1482 lines}). The enhanced
line emission in the forward facing component of the limb brightened
bi-cone is clearly visible above the disk (particularly towards the
bottom of the \Nii\ and \Oiii maps of Fig.~\ref{ngc1482 lines}) with
the counter cone visible, subject to dust attenuation, below the
inclined disk in most emission lines.  Line profiles from the near and
far side of the extended filaments show clear evidence of line
splitting at the level of $\sim$130\kms.

Strong line emission is seen throughout the galactic disk with the
base of the outflow cone coincident with the outer boundary of strong
star formation (as measured by the gradient in the \Ha\ emission along
the disk), indicative of a wind launched above (and below) the plane
of the disk from an extended starburst region.  A sharp gradient in
the electron density, as measured by the \Sii:\Sii\ ratio
(Fig.~\ref{ngc1482 ratios}) is also coincident with the boundaries of
the base of the outflow region, suggestive of a clearing of the inner
disk by the action of the starburst wind.

Under an alternate intensity scaling, the \Ha\ integrated flux map
does show a compact nuclear region at the limit of the spatial
resolution of the SPIRAL data. The nucleus appears to be resolved at
mid-IR wavelengths \citep{Siebenmorgen08} and with no spectroscopic
evidence of an AGN the emission is likely that of a compact star
forming region.

The line ratio diagnostics for NGC~1482 are shown in Fig.~\ref{ngc1482
VO}. The limited sensitivity to \Oiii\ emission, particularly in the
counter flow region to the north of the disk (presumably hampered by
significant dust attenuation) restricts the region that can be
analysed in this manner.  As for all of our sample, most of the data
points are bunched to the right of the IDD, consistent with metal
enriched gas in the nuclear regions of massive galaxies.  The body of
the disk emission is consistent with high metallicity star formation
(Z$>$1.0, \citet{kew2001}).

The emission from the bi-cone of NGC~1482 is compatible with
shock-induced ionisation \citep{vr2002}. The line ratios \emph{harden}
with increasing height above the disk place.  This could indicate the
gas becomes increasingly shocked at greater distance above the disk
(not unexpected since the wind speed will increase with reduced ISM
gas density \citep{Rand98}), or may simply be a product of dilution of
the outflow emission line components blended with the underlying disk
emission in the simple single Gaussian line apporoximation currently
implemented. The line ratio maps indicate harder line ratios on the
western side of the southern lobe, coincident with the x-ray emitting
gas presented by \citet{Strickland2004b}.

\smallskip
\subsubsection*{NGC~253}
Due to its proximity, NGC~253 is one of the best studied examples of
the class of luminous infrared galaxies known to be undergoing an
intense burst of star formation \citep{Rieke80}.  \citet{Demoulin70}
interpreted the disturbed kinematics of the emission line gas as a
conic outflow which was later shown \citep{McCarthy87} to be
qualitatively consistent within the \citet{Chevalier85} starburst wind
model.  Evidence for entrainment of molecular gas within the outflow
has been reported \citep{Sakamoto06}.  The outflow is clearly visible
in the UV although the observed flux is most likely light from the
central starburst scattered by dust entrained in the outflow
\citep{Hoopes05}.  X-ray observations do not reveal evidence for an
AGN, being consistent with a starburst driven galactic wind
\citep{Persic98,Strickland2000}.

\begin{figure*}
\begin{center}
\includegraphics[width=80mm]{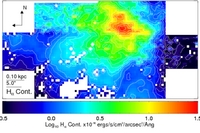}	    
\includegraphics[width=80mm]{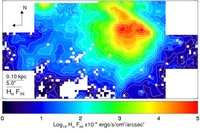}
\includegraphics[width=80mm]{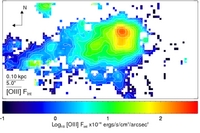}
\includegraphics[width=80mm]{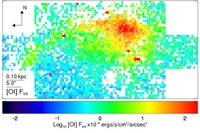}\\
\includegraphics[width=80mm]{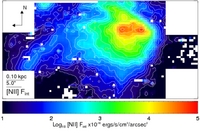}		 
\includegraphics[width=80mm]{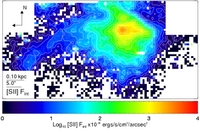}		 
\end{center}
  \caption{\label{ngc253 lines} NGC~253: Emission line maps are generated
  as for Fig.~\ref{ngc1482 lines}.}
\end{figure*}

\begin{figure}
\begin{center}
\includegraphics[width=80mm]{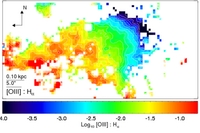} 
\includegraphics[width=80mm]{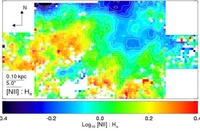}
\includegraphics[width=80mm]{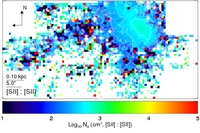}
\includegraphics[width=80mm]{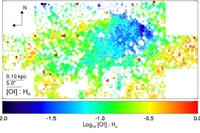} 
\includegraphics[width=80mm]{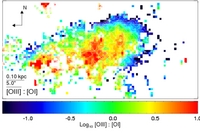} 
\end{center}
  \caption{\label{ngc253 ratios} NGC~253: Line ratio maps are
  constructed following Fig.~\ref{ngc1482 ratios}.}
\end{figure}

\begin{figure*}
\begin{center}
\includegraphics[width=50mm]{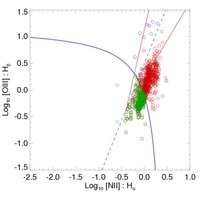}
\includegraphics[width=50mm]{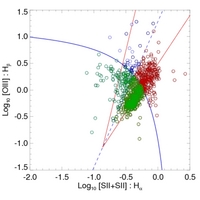}
\includegraphics[width=50mm]{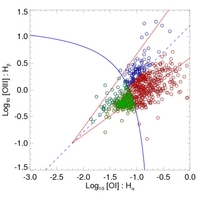}\\
\includegraphics[width=50mm]{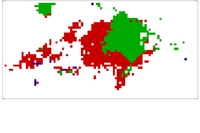}
\includegraphics[width=50mm]{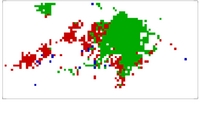}
\includegraphics[width=50mm]{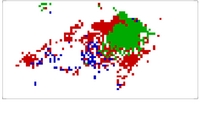}
\end{center}
  \caption{\label{ngc253 VO} NGC~253: Ionisation Diagnostic Diagrams
  are constructed following Fig.~\ref{ngc1482 VO} and using the same
  fiducial indicators.}
\end{figure*}

The large spatial extent of NGC~253 meant that only the southern side
of the galactic plane could be surveyed with the SPIRAL IFU in the
allocated observing time ($\sim$4 hours per object).  The observations
are presented in Figs.~\ref{ngc253 lines}$-$\ref{ngc253 VO}.  NGC~253
closely follows the emission pattern of NGC~1482.  Two prominent
fingers of emission, interpreted as the limb brightened edges of a
line-emitting cone, are clearly identified in the SPIRAL data
(Figs.~\ref{ngc253 lines}), with the structure most clearly
highlighted by the contours of \Nii\ emission. Like M82, the \Oiii\
emission is more centrally condensed than the low ionisation emission
(cf.\ Shopbell \& Bland-Hawthorn 1998).  Much like in NGC~1482, the
prominence of the \Nii\ emission is suggestive of shock excitation at
the boundary of the flow.

Line splitting, indicative of a hollow conic outflow structure, is
seen in spectra taken from the central region of the outflow above the
plane of the disk. The velocity separation is roughly $\sim$100\kms\
close to the disk but rises to $>$230\kms\ at the edge of the IFU
mosaic suggesting an acceleration of the outflow with increasing
height above the disk plane (under the assumption of a constant cone
opening angle). The limb brightened cone edges show only a single
velocity component.  The wind filaments converge to a considerably
more compact region ($\sim 140$ pc) in the nuclear disk compared to
what is seen in NGC~1482.

Consistent with \citet{Strickland2000}, the western limb is brighter
in the Balmer lines. In contrast, we find that the eastern limb is
consistently brighter in all forbidden lines relative to \Ha\
(Fig.~\ref{ngc253 ratios}). A similar asymmetry is seen in NGC~1482.
We observe an apparent hardening of the line ratios with height above
the disk, but this may be due to dilution from the inner disk.

The IDDs constructed from the SPIRAL data (Fig.~\ref{ngc253 VO}) reveal
the dominance of shock ionisation in the optical filaments. In particular, 
weak \Oi\ emission is observed across
the cone being especially prominent in the eastern limb. At large distances,
the weaker \Oi\ emission is compromised by the proximity of strong 
telluric \Oi\ emission due to the low redshift of NGC~253.

Consistent with this picture, \citet{Strickland2000} conclude that the
outflow emission seen at x-ray energies is derived from a low filling
factor gas that is excited by interaction between the wind and the ISM
on the surface of the outflow cone, rather than from the wind itself.
In conclusion, NGC~253 exhibits a starburst driven, shock excited wind
nebula, observed in this instance only on the near side of the disk
due to the spatial coverage of the data.  The wind originates from a
centrally confined starburst, rather than from a significant portion
of the disk region.

\smallskip
\subsubsection*{NGC~1808}
The ``peculiar'' nuclear spectrum of this object led
\citet{Veron85} to classify NGC~1808 as a weak Seyfert 2 galaxy.
Subsequent observations \citep{Forbes92} showed that the properties
of the nuclear emission are more plausibly explained by a nuclear
starburst, a view supported by mid-IR diagnostics \citep{Laurent00}.
\citet{Phillips93} confirm the nuclear outflow suspected
from earlier work, arguing against the Seyfert 2 identification
in favour of a starburst-driven outflow source.

\citet{Junkes95} find that the nuclear x-ray emission from the
unresolved core is largely due to the SNRs and hot bubbles of the
central starburst. Near-IR spectroscopy of the nucleus
\citep{Krabbe94} favours a relatively young starburst (age $\sim$
50\,Myr with a star formation rate of $\sim$1 M$_{\odot}$ yr$^{-1}$);
see \S 4.1). But at higher energies, there is evidence for a heavily
obscured AGN component (q.v. \citet{xmm1808}).
Thus, we retain the Seyfert 2 classification in Table 2 for this
galaxy, although the designation remains uncertain.

\begin{figure*}
\begin{center}
\includegraphics[width=70mm]{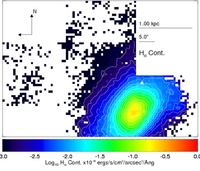}		    
\includegraphics[width=70mm]{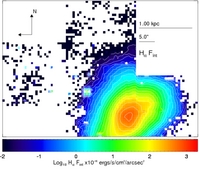}\\		    
\includegraphics[width=70mm]{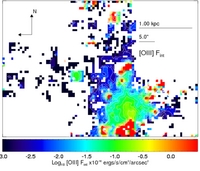}\\
\includegraphics[width=70mm]{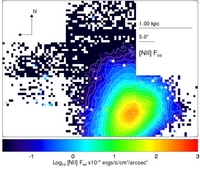}	    
\includegraphics[width=70mm]{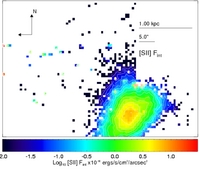}	    
\end{center}
  \caption{\label{ngc1808 lines} NGC~1808: Emission line maps are
  generated as for Fig.~\ref{ngc1482 lines}. Observations of NGC~1808
  were undertaken during a period of poor atmospheric
  transparency and so emission line maps suffer reduced sensitivity.
  No \Oi\ emission was recovered.}
\end{figure*}

\begin{figure}
\begin{center}
\includegraphics[width=70mm]{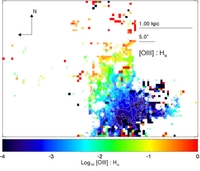}
\includegraphics[width=70mm]{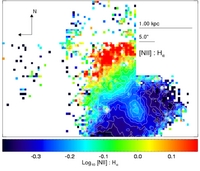}
\includegraphics[width=70mm]{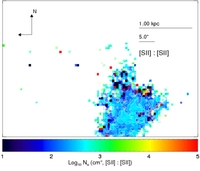}
\end{center}
  \caption{\label{ngc1808 ratios} NGC~1808: Line ratio maps are
  constructed following Fig.~\ref{ngc1482 ratios}. Observations of
  NGC~1808 were undertaken during a period of poor atmospheric
  transparency and so emission line maps suffer reduced sensitivity.}
\end{figure}

\begin{figure*}
\begin{center}
\includegraphics[width=55mm]{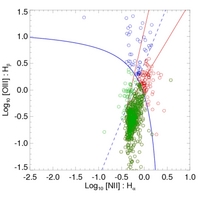}
\includegraphics[width=55mm]{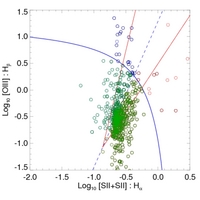}\\
\includegraphics[width=55mm]{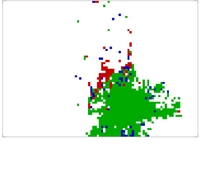}
\includegraphics[width=55mm]{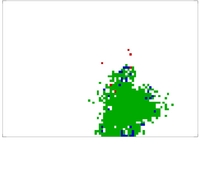}
\end{center}
  \caption{\label{ngc1808 VO} NGC~1808: Ionisation Diagnostic Diagrams
  are constructed following Fig.~\ref{ngc1482 VO}. No \Oi\ emission was
  recovred and so the third IDD cannot be constructed.}
\end{figure*}

Figs.~\ref{ngc1808 lines}$-$\ref{ngc1808 VO} present the new SPIRAL
observations.  NGC~1808 was observed under less than ideal conditions,
with variable seeing and intermittent cloud cover.  But we were able
to extract useful IDDs for the inner extent of the wind.  Our
observations reproduce the asymmetric line profiles of
\citet{Phillips93}, with blue and red shoulders corresponding to
the near side outflow and counter flow, as previously reported.  Due to
the weakness of \Oiii\ and \Oi, the IDDs do not provide an unambiguous
answer for the nature of the ionizing source in the extended
filaments. But we stress that the base of the wind is enhanced in
\Nii/\Ha, similar to what is observed in NGC~1482, and the line
profiles are somewhat broadened. These observations are consistent
with large-scale shocks in the outflowing gas.

\smallskip
\subsubsection*{NGC~3628}
\citet{Fabbiano99} presented the first strong
evidence from x-ray and \Ha\ observations for a collimated outflow
suspected from earlier kinematic work
\citep{Schmelz87a,Schmelz87b}. Subsequently, a corresponding molecular
disk-halo outflow was reported by \citet{Irwin96}. IR \citep{iras} and
radio \citep{Condon82} measurements place this object on the well
known far-IR vs.\ radio correlation indicating that the central
energetics are dominated by star formation. At x-ray energies, no
evidence is found for an active nucleus at the current detection
limits \citep{Flohic06}.

\begin{figure*}
\begin{center}
\includegraphics[width=70mm]{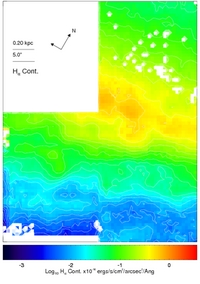}		    
\includegraphics[width=70mm]{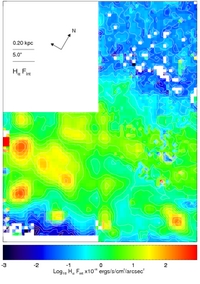}\\		    
\includegraphics[width=70mm]{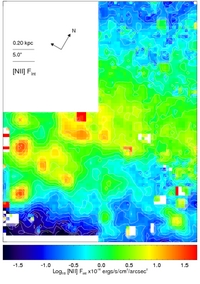}   
\includegraphics[width=70mm]{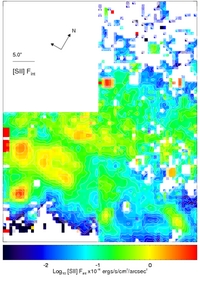}	    
\end{center}
  \caption{\label{ngc3628 lines} NGC~3628: Emission line intensity maps
  following Fig.~\ref{ngc1482 lines}.  Observations of NGC~3628 were
  undertaken during a period of poor atmospheric transparency and so
  emission line maps suffer reduced sensitivity.  No \Oi\ emission was
  recovered.}
\end{figure*}

\begin{figure}
\begin{center}
\includegraphics[width=70mm]{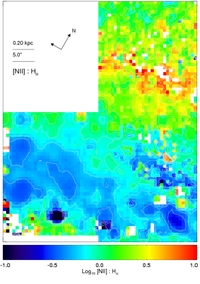}
\end{center}
  \caption{\label{ngc3628 ratios} NGC~3628: Line ratio maps are
  constructed following Fig.~\ref{ngc1482 ratios}. Due to the reduce
  sensitivity of the observations for NGC~3628, only the \Nii/\Ha\ map
  is recovered.}
\end{figure}

Observations of NGC~3628 were undertaken in less than ideal observing
conditions. We were able to extract useful spectral diagnostic
information from the red data, but only limited information from the
blue data. Emission line maps were extracted for \Ha, \Nii\ and \Sii\
and are presented in Figs.~\ref{ngc3628 lines} \& \ref{ngc3628
ratios}.  The strongest line emission originates in a series of
compact regions to the south-east of the SPIRAL mosaic, which we
interpret as \HII\ regions. The \Nii/\Ha\ ratio image
(Fig.~\ref{ngc3628 ratios}) reveals the filamentary emission above the
plane of the disk reported by \citet{Fabbiano99}.

The line ratios in this region show enhanced \Nii/\Ha\ similar to what
is observed in most of our starburst wind sources. Strong \Oiii/\Hb\ 
is ruled out by the blue data even with its compromised sensitivity.
The base of the wind is
enhanced in \Nii/\Ha, reminiscent of what is observed in NGC~1482, 
and the line profiles are somewhat broader than line
emission from neighbouring HII regions. Once again, these observations
are consistent with large-scale shocks in the outflowing gas.

\smallskip
\subsubsection*{NGC~6810}
\begin{figure*}
\begin{center}
\includegraphics[width=70mm]{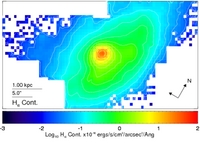}	    
\includegraphics[width=70mm]{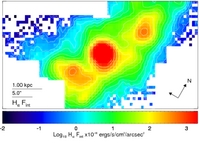}\\
\includegraphics[width=70mm]{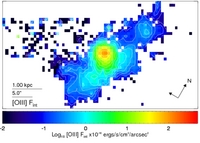}\\
\includegraphics[width=70mm]{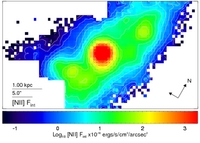}		 
\includegraphics[width=70mm]{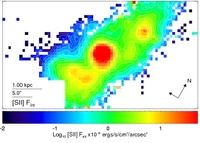}		 
\end{center}
  \caption{\label{ngc6810 lines} NGC~6810: Emission line intensity maps
  following Fig.~\ref{ngc1482 lines}. The \Oiii\ map is limited in extent
  by the sensitivity of the data, and no \Oi\ map was recovered.}
\end{figure*}

\begin{figure*}
\begin{center}
\includegraphics[width=70mm]{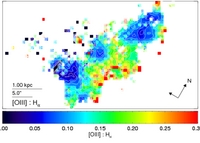}
\includegraphics[width=70mm]{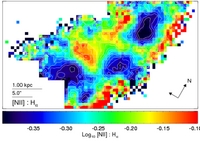}\\ 
\includegraphics[width=70mm]{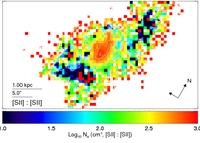}
\end{center}
  \caption{\label{ngc6810 ratios} NGC~6810: Line ratio maps are
  constructed following Fig.~\ref{ngc1482 ratios}.}
\end{figure*}

\begin{figure*}
\begin{center}
\includegraphics[width=55mm]{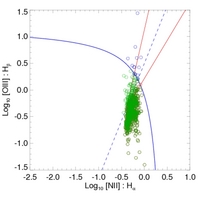}
\includegraphics[width=55mm]{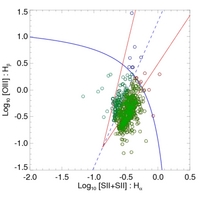}\\
\includegraphics[width=55mm]{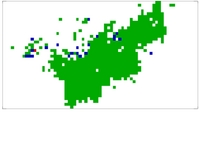}
\includegraphics[width=55mm]{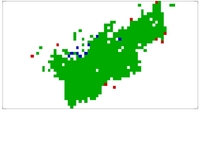}\\
\end{center}
  \caption{\label{ngc6810 VO} NGC~6810: Ionisation Diagnostic Diagrams
  are constructed following Fig.~\ref{ngc1482 VO}. No \Oi\ emission was
  recovred and so the third IDD cannot be constructed for NGC~6810.}
\end{figure*}

The historic classification of this object
as a Seyfert 2 galaxy is based primarily on the width of emission
lines in low spatial resolution spectra of the nuclear region.
\citet{Coccato04} noted ``a higher than expected minor axis velocity
dispersion and unusual kinematics'' in long slit spectroscopy.
\citet{str2007} concluded that NGC~6810 is in fact a ``full disk''
superwind galaxy and that the previous classification is based largely
on misidentification of counterparts in a low resolution radio survey
and the merged line profiles of an outflowing wind.  This is
compatible with the earlier observation \citep{Forbes98} that the
radio, FIR and [Fe\,\textsc{ii}] properties of the inner disk are
consistent with star formation.  Our new SPIRAL observations
(Figs.~\ref{ngc6810 lines}$-$\ref{ngc6810 VO}) are consistent with
this interpretation. There is no evidence for AGN activity in the
optical spectra, with narrow lines consistent with solar-metallicity
star formation across the stellar disk.  Star formation across the
disk of the galaxy dominates the IDDs.

The \Oiii/\Ha\ line ratio map (Fig.~\ref{ngc6810 ratios}) shows
striking evidence for an ``ionisation cone'' on the near side of the
disk. These structures are not unique to AGN sources (e.g.\ Pogge 1987;
Tadhunter \& Tsvetanov 1989) having been observed in starburst
galaxies (e.g.\ Shopbell \& Bland-Hawthorn 1998). Leaked radiation from
a starburst or an AGN, or even shocks excited by a conic outflow, can
produce ionisation cones along the minor axis. In support of this
statement, the \Oiii/\Hb\ ratio in this region is characteristically
low (cf. \citet{shu1979}).  In the line ratio map, we have used \Ha\ in 
place of \Hb\ because the stronger \Ha\ signal reveals the cone more
clearly.

Unfortunately, the weakness of \Oiii\ emission above the galactic disk
makes it difficult to unambiguously classify the excitation mechanism
for the outflow gas from the IDD (Fig.~\ref{ngc6810 VO}). But we
stress that the base of the wind is enhanced in \Nii/\Ha\ similar to
what is observed in NGC~1482. The high \Nii/\Ha\ gas delineates the
boundary of a chimney structure emanating from the central starburst
region. The base of the chimney extends $\sim$1.5~kpc across the inner
galaxy. An inspection of emission-line profiles in regions of enhanced
\Nii/\Ha\ reveals multi-component line profiles indicative of high
velocity material in a blueshifted outflow on the eastern side of the
disk, with the corresponding counter flow redshifted (and dust
attenuated) on the opposite side. In summary, our observations provide
further evidence that the off-nuclear emission in NGC 6810 is due to a
large-scale galactic wind. Like NGC~1482, the dominant source of
ionisation is most likely due to shock excitation of filaments
entrained in the wind.

\subsection{AGN}
\label{AGN}

\smallskip
\subsubsection*{NGC 1365}
Optical spectroscopy of this well known Seyfert
\citep{Veron80,Edmunds82} reveals a broad component to the \Ha\ line
and line ratios indicative of AGN activity. Further evidence for the
AGN comes from x-ray variability of the highly obscured nucleus
\citep{Ward78}.  But based on an analysis of the radio and x-ray
structure of the central regions, \citet{Stevens99} suggest that the
energetics are dominated by star formation, and that ``the evidence of
a jet emanating from the nucleus is at best marginal'' (cf.\
\citet{Sandqvist95}).  The radio/far-IR ratio is also indicative of a
dominant circumnuclear starburst \citep{Forbes98}.

The extended ionised gas in NGC~1365 has been extensively studied 
with long-slit spectroscopy. On the basis of these observations,
\citet{Hjelm96} propose a physical structure (wide angle, hollow conic
outflow) to explain the unusual velocity structures previously
reported \citep{Phillips83,Jorsater84} in the context of an outflow
(first suggested by \citet{Burbidge60}), while
\citet{Storchi-Bergmann91} present clear evidence of an \Oiii\
``ionisation cone'' to the south east of the nucleus, above the plane of
the disk.

\begin{figure*}
\begin{center}
  \includegraphics[width=70mm]{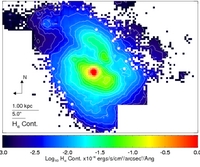}
  \includegraphics[width=70mm]{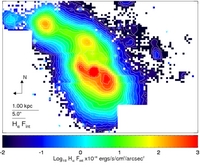}\\
  \includegraphics[width=70mm]{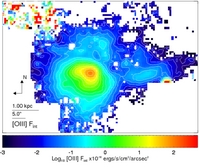}
  \includegraphics[width=70mm]{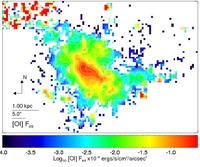}\\
  \includegraphics[width=70mm]{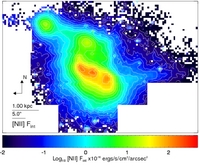}
  \includegraphics[width=70mm]{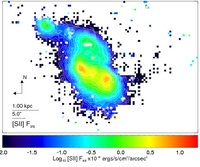}
\end{center}
  \caption{\label{ngc1365 lines} NGC~1365: Emission line intensity maps
  following Fig.~\ref{ngc1482 lines}.}
\end{figure*}

\begin{figure*}
\begin{center}
  \includegraphics[width=70mm]{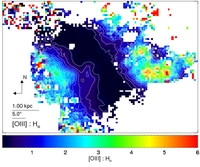}
  \includegraphics[width=70mm]{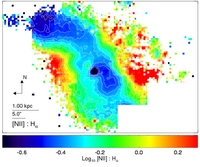}\\
  \includegraphics[width=70mm]{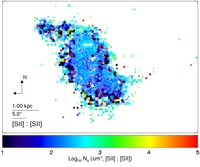}
  \includegraphics[width=70mm]{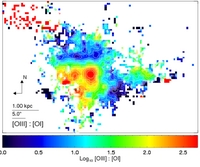}
\end{center}
  \caption{\label{ngc1365 ratios} NGC~1365: Line ratio maps are
  constructed following Fig.~\ref{ngc1482 ratios}.}
\end{figure*}

\begin{figure*}
\begin{center}
\includegraphics[width=50mm]{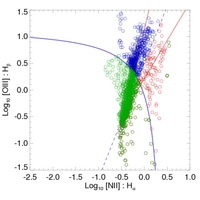}
\includegraphics[width=50mm]{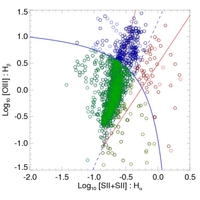}
\includegraphics[width=50mm]{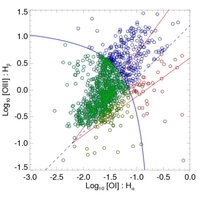}\\
\includegraphics[width=50mm]{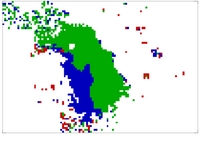}
\includegraphics[width=50mm]{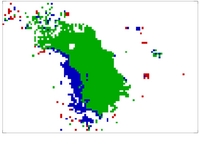}
\includegraphics[width=50mm]{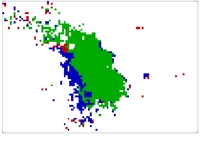}
\end{center}
  \caption{\label{ngc1365 VO} NGC~1365: Ionisation Diagnostic
  Diagrams are constructed following Fig.~\ref{ngc1482 VO}.}
\end{figure*}

The new SPIRAL observations for NGC~1365 are presented in
Figs.~\ref{ngc1365 lines}$-$\ref{ngc1365 VO}. The reconstructed continuum
image reproduces the central disk structure seen in the
{\it HST} imaging from Fig.~\ref{finding charts}. The \Ha\ integrated
flux map suggests strong star formation in the nuclear zone, with a
number of probable compact \HII\ regions or star clusters.
  
Strong \Oiii\ emission is seen to extend eastwards from the nuclear
region above the plane of the disk, while a counterpart emission
region seen below the disk is at lower intensity, presumably due to
dust attenuation on crossing the plane of the nuclear disk.  \Nii\
emission exhibits the classic limb-brightened arms of a conic
structure to the west but such a feature is less obvious on the near
side of the disk, possibly due to confusion with disk emission.
\citet{Phillips83} report the ``remarkable kinematics of the ionised
gas in the nucleus of NGC 1365'' and line splitting to be most
prominent in the south-eastern region, particularly in the \Oiii\
line, observations supported by the SPIRAL data.  Line splitting is
evident but not complete at the resolution of the SPIRAL observations,
suggesting a narrow opening angle or low velocities for the outflow.

The emission-line ratio maps shown in Fig.~\ref{ngc1365 ratios} are
revealing.  The counter fan of \Oiii\ emission is evident to the west
in the \Oiii/\Ha\ ratio, while the enhanced \Nii/\Ha\ ratio usually
observed in outflowing emission line gas is seen to fill the region
bounded by the limb brightened \Nii\ structure of the western region
of Fig.~\ref{ngc1365 lines}.  The expected counter-flow region is
visible to the east as high \Nii/\Ha\ gas once far enough above the
disk to avoid contamination by the strong star formation in the
region.  The \Sii\ electron density diagnostic shows no clear
structure at the limit of our signal-to-noise limit.  The ionisation
cone from the central AGN is most evident on the nearside of the disk
in the \Oiii/\Oi\ ratio (Fig.~\ref{ngc1365 ratios}).

The classic IDDs are presented in Fig.~\ref{ngc1365 VO}: the dominant
signature is that of a high metallicity starburst consistent with
dominant circumnuclear star formation. However, significant emission,
particularly in the outflow regions, lies above the \emph{extreme star
formation} boundary of \citet{kew2001}.  The zone occupied by this
gas, in all three of the classic diagnostic diagrams, is indicative of
the hard ionising spectrum expected from AGN photoionisation, and is
for the most part inconsistent with photoionisation by starlight or
shock ionisation models. Broad He\textsc{ii}\,$\lambda$4686 and
He\textsc{i}\,$\lambda$5876 emission are observed in the nuclear
spectrum, consistent with photoionisation by a hard ionizing source.

\smallskip
\subsubsection*{Circinus}
Located at low galactic latitude and with high foreground extinction
(A$_V$=1.5), this large nearby Seyfert galaxy was first identified by
\citet{Freeman77}.  Maser measurements firmly establish the central
black hole mass at $m_{\rm BH} \approx 2\times10^6$ M$_\odot$
\citep{Greenhill03a}.  With the inferred accretion rate ($\approx$0.1
$L_{\rm{edd}}$) at odds with much higher accretion rates in other
Seyferts with maser-determined black hole masses (e.g.\ NGC 1068;
Begelman \& Bland-Hawthorn 1997), Circinus lends support to the notion
that supermassive black holes are indeed quite common in galaxies, but
for the most part have long duty cycles and/or low accretion rates.

The Seyfert 2 nature of Circinus is supported by the mid-IR coronal
line emission \citep{Lutz02}, e.g.\ [Si\,\textsc{ix}] at 3.94$\mu$m
\citep{Oliva94}.  Recent integral field spectroscopy, assisted by
adaptive optics \citep{sinfoniCircinus06}, indicates a substantial
(1$-$2\%\,$L_{\rm bol}$) young ($<$100~Myr) toroidal starburst within
10~pc of the central AGN and provides further evidence for a ``wide
angle outflow'' as reported by \citet{Greenhill03b}.  Extended
filamentation associated with the outflow was first mapped by
\citet{vb1997}, and later in more detail by \citet{ttf03}.  X-ray
spectral analysis further supports the view of Circinus as a highly
obscured AGN but with strong thermal dust emission dominated by
heating from nuclear star formation \citep{Matt00}.

\begin{figure*}
\begin{center}
\includegraphics[width=70mm]{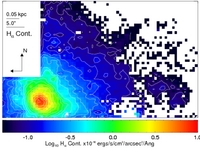}	    
\includegraphics[width=70mm]{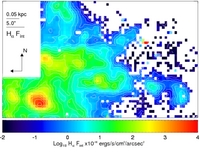}\\
\includegraphics[width=70mm]{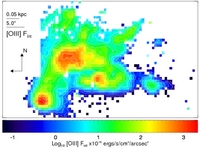}\\
\includegraphics[width=70mm]{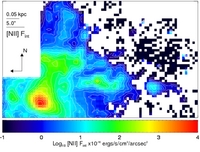}		 
\includegraphics[width=70mm]{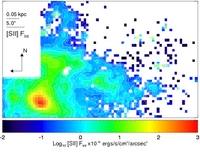}		 
\end{center}
  \caption{\label{circinus lines} Circinus: Emission line intensity maps
  following Fig.~\ref{ngc1482 lines}.}
\end{figure*}

\begin{figure*}
\begin{center}
\includegraphics[width=70mm]{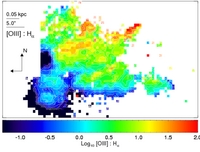}	 
\includegraphics[width=70mm]{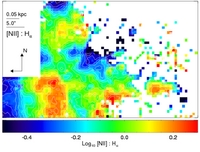}\\	 
\includegraphics[width=70mm]{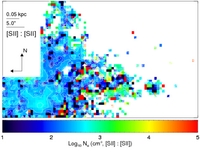}
\includegraphics[width=70mm]{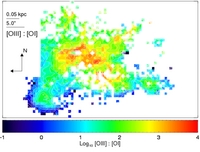}
\end{center}
  \caption{\label{circinus ratios} Circinus: Line ratio maps are
  constructed following Fig.~\ref{ngc1482 ratios}.}
\end{figure*}

\begin{figure*}
\begin{center}
\includegraphics[width=50mm]{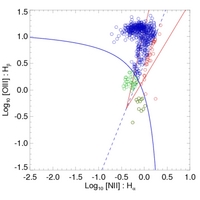}
\includegraphics[width=50mm]{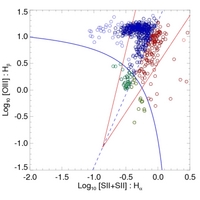}
\includegraphics[width=50mm]{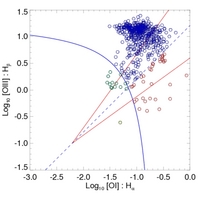}\\
\includegraphics[width=50mm]{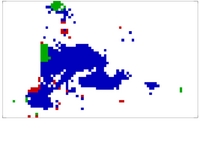}
\includegraphics[width=50mm]{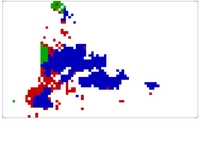}
\includegraphics[width=50mm]{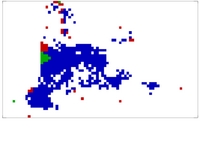}
\end{center}
  \caption{\label{circinus VO} Circinus: Ionisation Diagnostic
  Diagrams are constructed following Fig.~\ref{ngc1482 VO}.}
\end{figure*}

The large spatial extent of the Circinus galaxy allowed only the
north-west side of the disk to be probed with
SPIRAL. Figs.~\ref{circinus lines}$-$\ref{circinus VO} present the
observations.  Our spatial coverage was guided by \emph{TTF}
observations of \citet{vb1997} and the \emph{HST} study of
\citet{wss2000}.  We have not attempted to correct the observed
spectra for foreground reddening.  The ``hook'' structure identified
by \citet{vb1997} traces the southern boundary of an ionisation cone,
running east-west in the prominent emission lines.

The \Oiii\ line profiles show significant complexity with a number of
mixed velocity components. Maximal line splitting in the centre of the
ionisation cone, and single line profiles along its outer walls, might
suggest a simple conic outflow structure.  But the line intensity maps
indicate limb darkening rather than the limb brightening one would
expect on a conic emitting surface at the interface of a hot wind
material and the cooler ISM, suggesting significant emission from gas
within the cone.
There is a strong velocity gradient across the filaments, each
filament being at a different mean velocity, with the \Oiii\ emission
from the east-west boundary significantly redshifted with respect to
the opposite (north-south) region of the ionisation cone. The ouflow
is best interpreted as a gas-filled ionisation cone illuminated by the
central AGN, rather than a walled cavity structure.

\smallskip
\subsubsection*{NGC~5128 (Centaurus A)}
Cen A is one of the most well studied active galaxies owing to its
extraordinary brightness across 16 decades or more of the
electromagnetic spectrum (Israel 1990).  It is the nearest powerful
AGN and, more specifically, the closest FR~I radio
source. Fortuitously, we are witnessing an accretion-powered radio
source in the first 100~Myr of its projected $0.5-1$~Gyr lifetime.  A
well known feature of Cen A, unique among our sample, is the presence
of a spectacular radio jet \citep{Blanco75,Peterson75} which ignites
optical filaments along its axis \citep{Sutherland93}. The jet-induced
emission falls far beyond the radial extent of our IFU mosaic.

Over the inner galaxy, the extended ionised gas was first mapped by
\citet{Bland87} and \citet{Nicholson92}, and shown to arise
exclusively from a severely-warped thin disk, later confirmed by CO
observations \citep{Quillen93}. A diffuse highly ionised halo of
\Oiii\ was first identified by Bland-Hawthorn \& Kedziora-Chudczer
(2003, see their Fig.~6) and has recently been shown to match the
extended x-ray maps from {\emph Chandra} \citep{Kraft08}
extraordinarily closely.  Extended mid-IR emission is also seen along
the radio axis, revealing evidence for a limb-brightened shell
\citep{Quillen06}, and providing strong evidence for a bipolar outflow
along the radio axis.

\begin{figure*}
\begin{center}
\includegraphics[width=50mm]{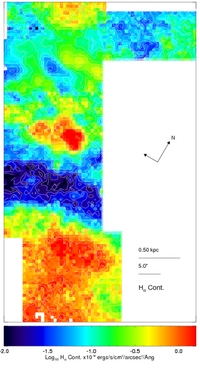}	    
\includegraphics[width=50mm]{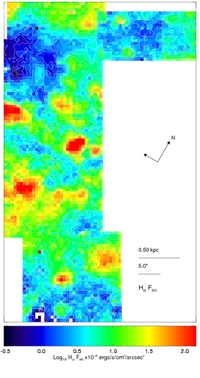}
\includegraphics[width=50mm]{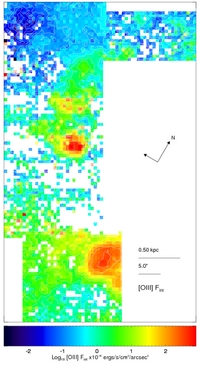}\\
\includegraphics[width=50mm]{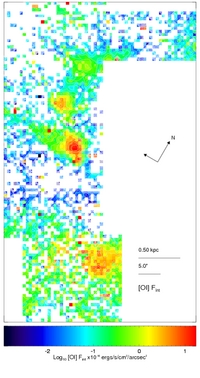}
\includegraphics[width=50mm]{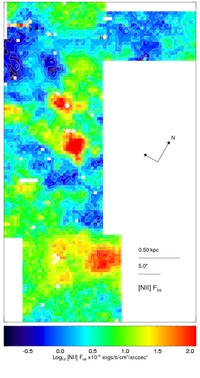}		 
\includegraphics[width=50mm]{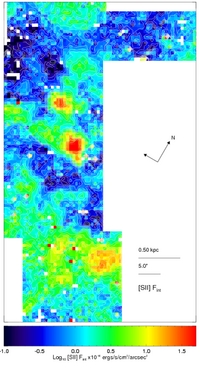}		 
\end{center}
  \caption{\label{CenA lines} NGC~5128 (CenA): Emission line intensity
  maps following Fig.~\ref{ngc1482 lines}.}
\end{figure*}

\begin{figure*}
\begin{center}
\includegraphics[width=60mm]{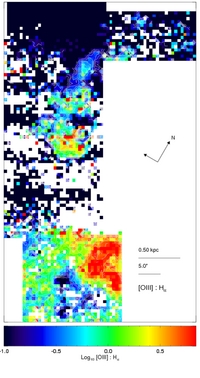}	 
\includegraphics[width=60mm]{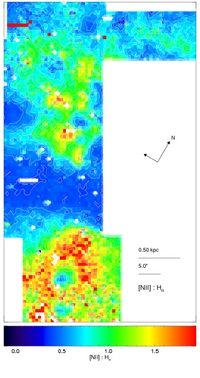}\\	 
\includegraphics[width=60mm]{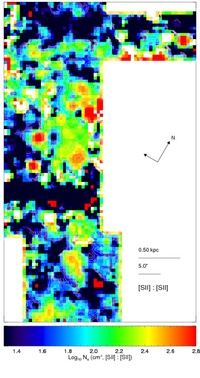}
\includegraphics[width=60mm]{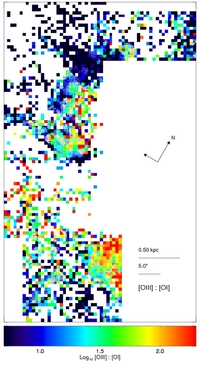}
\end{center}
  \caption{\label{CenA ratios} NGC~5128 (CenA): Line ratio maps are
  constructed following Fig.~\ref{ngc1482 ratios}. Note the data cube
  was block averaged 3$\times$3 prior to spectral fitting in the two
  \Sii\ lines for the construction of the electron density diagnostic
  due to the weak signal.}
\end{figure*}

\begin{figure*}
\begin{center}
\includegraphics[width=50mm]{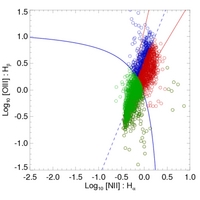}
\includegraphics[width=50mm]{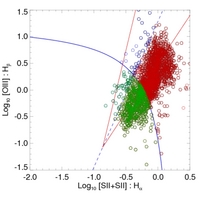}
\includegraphics[width=50mm]{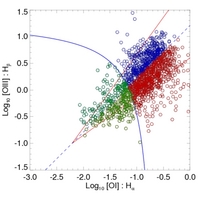}\\
\includegraphics[width=50mm]{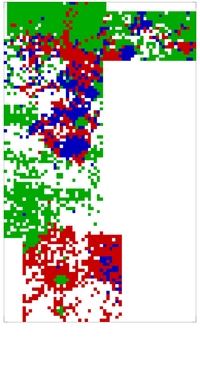}
\includegraphics[width=50mm]{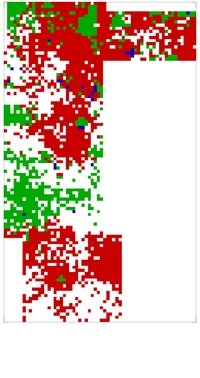}
\includegraphics[width=50mm]{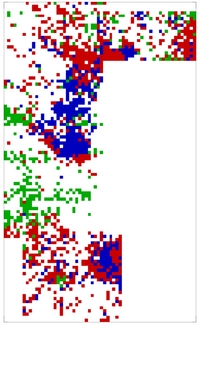}
\end{center}
  \caption{\label{CenA VO} NGC~5128 (CenA): Ionisation Diagnostic
  Diagrams are constructed following Fig.~\ref{ngc1482 VO}.}
\end{figure*}

\begin{figure*}
\begin{center}
\includegraphics[width=85mm]{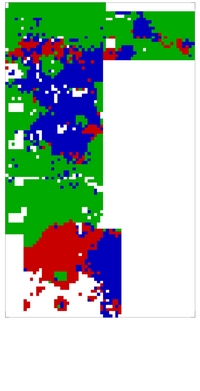}
\end{center}
  \caption{\label{CenA cone} NGC~5128 (CenA): The ionisation cone due
  the hard AGN spectrum is revealed more clearly in the data after
  spatially re-binning the spectra 3$\times$3.  The classification
  scheme from the \Oiii/\Hb\ vs.\ \Nii/\Ha\ IDD (Fig.~\ref{CenA VO})
  is repeated here.  }
\end{figure*}

Here we provide only a brief summary of the SPIRAL observations
presented in Figs.~\ref{CenA lines}$-$\ref{CenA VO}.  A more detailed
discussion and associated modeling is to be presented elsewhere.  The
\Oiii/\Hb\ vs.\ \Nii/\Ha\ IDD reveals a large number of data points
above the extreme starburst line.  When these points are identified in
the SPIRAL data, they reveal a striking nuclear ionisation cone.  The
cones vissibility is enhanced after a $3\times3$ spatial binning of
the mosaiced blue IFU data (Fig.~\ref{CenA cone}).  The conic region,
delineated by sharp edges, exhibits gas with the ionisation ratios
expected from a hard AGN ionising spectrum, although shocks cannot be
ruled out. The structure can be traced back towards the nucleus where
it likely connects with the Pa$\beta$ emission cone seen by
\citet{krajnovic2007} at near-IR wavelengths.  The feature is roughly
aligned with the known radio jet structure above the plane of the
disk.  Our coverage to the south is insufficient to trace a possible
counter cone. Throughout the cone, we clearly see variations in the
\Sii\ doublet consistent with density variations of order $n_e \approx
300-1000$ cm$^{-3}$. This is consistent with the inferred
shell-enhanced densities from the mid-IR observations
\citep{Quillen06}.

In summary, we have discovered an ``ionisation cone'' along the minor
axis of Centaurus A for the first time. The phenomenon is most evident
in emission-line ratio images (e.g.\ \Oiii/\Hb) where a clear
distinction can be made between the ionising source in the outflow
from the background ionising source. This tends to favour an
ionisation cone powered by an AGN source, but the phenomenon can also
be identified in starburst galaxies (see \S 4.1), as we discovered
here in NGC~6810.

\smallskip
\subsubsection*{IC~5063}
\citet{Bergeron83} presented early evidence for an obscured Seyfert
nucleus via the detection of a broad yet weak component to the \Ha\
emission. \citet{Colina1991} reported high-excitation emission lines
and confirmed the broad \Ha\ profile, thereby supporting the claim of
a concealed active nucleus.  Both papers note the kinematic decoupling
of gas in the inner regions from the extended disk rotation curve of
the galaxy.  \citet{Colina1991} favoured an explanation invoking
outflowing gas, and provided evidence of a wide angle (50$^\circ$)
ionisation cone via the spatial map of the \Oiii/\Ha\ ratio.

At radio wavelengths, \citet{mor2007} present high quality observations
of warm and cool material moving outwards from the nucleus (see also
\citet{morganti98}).  Their observations suggest a radio lobe/jet structure which
in turn may be responsible for the disturbed kinematics of the outflow gas.
These authors conclude there is little evidence for shock-driven excitation,
proposing instead that the emission is overwhelmed by AGN photoionisation.

The SPIRAL observations are shown in Figs.~\ref{ic5063
lines}$-$\ref{ic5063 VO}.  As presented in previous work
\citep{Bergeron83,Colina1991}, the velocity field of the emission line
gas is dominated by a regular disk rotation beyond 2\arcsec\
from the nucleus, although line profiles show evidence of broadening
at the base of the stronger lines, and line splitting is apparent
above the mid-plane.  The kinematically distinct outflowing gas shows
little evidence of the classic line splitting expected from a hollow
conic outflow structure, indicative of a more homogeneous gas
distribution in this region.  The ionisation diagnostic diagrams
(Fig.~\ref{ic5063 VO}) indicate AGN photoionisation almost
exclusively, at least in regions where the intensity of emission is sufficently
above the fitting threshold.

\smallskip
\subsubsection*{NGC~6240}
As one of the closest of the ultraluminous infrared galaxies (ULIRG),
NGC~6240 is central to the longstanding debate as to the presence or
absence of quasar-like nuclei in ULIRG sources.  In the case of
NGC~6240, hard x-ray observations unambiguously identify a double AGN
core at its heart \citep{Komossa2003}.  A large-scale wind was first
reported by Heckman, Armus and Miley (1990) (cf.\ \citet{Ptak03},
\citet{Colbert94}).  The extended emission-line gas has been mapped in
optical lines by imaging tunable filters \citep{jbh91,ttf03} and in
x-ray emission \citep{Komossa2003}.

The SPIRAL observations show that the nuclear regions of NGC~6240 are
highly complex both in the kinematics and ionised gas
distribution. With limited spatial coverage possible due to observing
constraints, our observations were guided by the study of
\citet{ttf03} and focus on the region of high \Nii/\Ha\ ratio
emission.  Two dimensional emission maps are not shown as the
complexity of the emission structure in the nuclear region combined
with the low surface brightness of emission in the less complex outer
zones is not easily interpreted in this manner.  Spectra from the
nucleus show kinematically broadened line emission (FWHM$\sim$13\AA)
with a velocity spread of 500-600\,\kms.  Most spectra do not easily
separate into identifiable \Ha\ and \Nii\ lines.

An emission-line filament is observed to extend east from the nucleus.
We interpret the isolated line in the filament spectrum as \Ha\ offset
in velocity from the host galaxy by $+340$\,\kms in the outer reaches
of the mosaic.
This is at odds with the \Ha\ vs.\ \Nii\ map of Veilleux\etal\ (2003)
but we note that the proposed shift in the emission wavelength could
lead to cross-contamination between the wavelength settings of the
tunable filter observations.  The conspicuous absence of satellite \Ha\
(and the fainter \Nii\ line) in the filament spectrum support our
proposed identification.
On heavily binning adjacent spectra, there is evidence for a weak
feature compatible with broad \Nii\ emission to accompany \Ha.  There
is tentative evidence for \Oiii\ emission associated with the \Ha\
filament as well, although the sensitivity of the data is not
sufficient to confirm this in detail.  In summary, we are unable to
construct IDDs for this object with the present observations.

\section{AGN vs. Starburst}
\label{AGN SF discussion}

In the previous section, we saw consistent differences in the ionisation 
characteristics of outflowing gas associated with starburst and AGN sources.
We now present a simple model that largely explains these differences. 
Consequently, our model provides a key insight into the galactic wind mechanism
and the radiative timescales of the starburst and AGN phenomenon.

\subsection{Predicted and observed levels of photoionisation}

Our galaxy sample spans bolometric luminosities in the range $L_{\rm bol} = 1-10\times 10^{10}$ L$_
\odot$ and are distinguished by their nuclear starburst or AGN activity. All of the galaxies 
were selected on the basis of known minor-axis outflows. 
The new observations reveal that all of the AGN galaxies display IDDs typical of direct 
photoionisation by the AGN, whereas the starburst galaxies display IDDs typical of shock 
ionisation. How are we to understand this?

First we note that both nuclear starburst and active galaxies almost certainly have duty 
cycles such that
what we observe now is more representative of their peak activity rather than their 
quiescent phase.
We now show that {\it prima facie} the distinction between starburst and AGN outflow
IDDs is surprising when one considers the overall ionizing photon production in both 
classes of sources.

A black hole that converts rest-mass energy with an efficiency
$\epsilon$ into radiation has a luminosity
\begin{eqnarray}
L_A &=& \epsilon \dot m c^2 \\
                     &=& 7\times 10^{11}\left({\epsilon\over 0.05}\right)\left({\dot m}\over {M_
\odot\; {\rm yr}^{-1}}\right)\;L_\odot
\end{eqnarray}
for which $\dot m$ is the mass accretion rate. In reality, the
accretion disk luminosity can limit the accretion rate through
radiation pressure. The so-called Eddington limit is given by
\begin{eqnarray}
L_E &=& {{4\pi G m_{\rm BH} m_p c}\over{\sigma_T}} \\
                      &=& 2\times 10^{11}\left({m_{\rm BH}\over 
10^7\;M_\odot}\right)\;L_\odot
\end{eqnarray}
where $m_{\rm BH}$ is the black-hole mass, $m_p$ is the proton mass
and $\sigma_T$ is the Thomson cross-section for electron
scattering. This equation allows us to derive a maximum accretion
rate $\dot m$, and therefore a maximum ionizing luminosity, for a given black-hole 
mass. We find that this relation is not strongly constrained with the present data.
For a canonical power-law
continuum ($\propto (h\nu)^{-1}$) typical of active galactic nuclei,
the number of Lyman continuum (LyC) photons is given by
\begin{equation}
{\cal N}_{\rm LyC,A} \sim 10^{54}\xi_A \left(L_{\rm bol,A} \over 10^{11} L_\odot \right)\; {\rm 
phot\ s}^{-1} 
\label{NLyCA}
\end{equation}
for which $\xi_A$ is the fraction of $L_{\rm bol}$ ($L_A = \xi_A L_{\rm bol}$) arising from
AGN activity today (i.e.\ not averaged over the lifetime of the source).

The total radiative output from a luminous starburst at the present epoch
can be approximated by (e.g.\ Maloney 1999; Heckman, Lehnert \& Armus 1993)
\begin{eqnarray}
L_S &\sim& 10^{11}\left({\dot s}\over {5\; M_\odot\; {\rm yr^{-1}}}\right)\left({\Delta
t\over 10^8\;{\rm yr}}\right)^{0.67} \\
       & \times & \left({m_L\over1\;M_\odot}\right)^{0.23} \left({m_U
\over 100\;M_\odot}\right)^{0.37}\;L_\odot
\end{eqnarray}
where $\Delta t$ is the starburst duration estimated from the gas
supply divided by the star formation rate $\dot s$. The
third and fourth terms indicate the lower and upper stellar mass
cut-offs. For a conventional IMF, the star formation rate is roughly
\begin{equation}
\dot s\sim 5\left({L_S} \over 10^{11}\;L_\odot\right)\left({\Delta t\over 10^8\;{\rm yr}}
\right)^{-0.67} \;M_\odot\; {\rm yr}^{-1} .
\end{equation}
We can now determine the rate of Lyman continuum photons produced by
the starburst. To a useful approximation,
\begin{eqnarray}
\label{eq:star}
{\cal N}_{\rm LyC,S} &\sim& 10^{54} \left({\dot s}\over{5\; M_\odot\; {\rm yr^{-1}}}
\right)\left({m_L\over 1\;M_\odot}\right)^{0.23} \left({m_U\over 100\;M_\odot}
\right)^{0.37} \\
             &\sim& 10^{54}\xi_S \left(L_{\rm bol} \over 10^{11} L_\odot \right)\; {\rm phot\ 
s}^{-1} 
\end{eqnarray}
which is to be compared to equation~\ref{NLyCA} above.  The
high-energy cut-off for the hot young stars declines rapidly beyond 50
eV. The factor $\xi_S$ is the fraction of $L_{\rm bol}$ ($L_S = \xi_S L_{\rm bol}$) arising 
from
the starburst activity today (i.e.\ not averaged over the lifetime of the source).

The LyC photon rate for both starbursts and AGNs for two sources with
the same bolometric luminosity is essentially the same.  While the
nonthermal spectrum of the AGN means that the harder photons propagate
at least an order of magnitude further into the surrounding gas, the
photon {\it number} is dominated by the soft photons so that the
comparison is still meaningful. We do not discuss the possibility of
the {\it total} accretion disk luminosity being channelled into a
smaller solid angle (i.e.\ intrinsically beamed) because we were unable
to find compelling evidence for this to be so.  This is to be
addressed in our later papers.

An ionizing photon rate as high as ${\cal N}_{\rm LyC}\sim 10^{54}\xi$
$(L_{\rm bol}/10^{11} L_\odot)$ phot s$^{-1}$ is easily large enough to account for the 
off-nuclear
line fluxes observed in all galaxies, whether starburst or AGN.
At a distance of 1 kpc, the ionizing flux is $\varphi \approx 10^{10}$ 
phot cm$^{-2}$ s$^{-1}$. We can relate this to the H$\alpha$ surface 
brightness through the emission measure ${\cal E}_m$ where
\begin{equation}
{\cal E}_m = \int f n_e^2\ {\rm d}z \ \ \ \ {\rm cm^{-6} \ pc}
\end{equation}
which is an integral of H recombinations along the line of
sight $z$ multiplied by a filling factor $f$.\footnote{${\cal E}_m(\rm H\alpha)=1$
cm$^{-6}$ pc is equivalent to 2$\times 10^{-18}$ erg cm$^{-2}$ s$^{-1}$
arcsec$^{-2}$ (cgs) for a plasma at 10$^4$ K.}
If we assume that the absorbing cloud along the minor axis is optically thick,
we can relate this directly to an H$\alpha$ surface brightness in cgs units
such that
\begin{equation}
\label{surfbri1}
\mu({\rm H\alpha}) = 1\times 10^{-14} \left({r}\over{{\rm 1\ kpc}}\right)^{-2} \left(L_{\rm 
bol} \over 10^{11} L_\odot \right)\ \ \ {\rm cgs}
\end{equation}
where $r$ is the distance to the central source
(Bland-Hawthorn\etal\ 1998, Appendix). Therefore
the expected H$\alpha$ surface brightness at 1 kpc is roughly $7\times 10^{-15}$ cgs 
for the AGNs, and $2\times 10^{-15}$ cgs for the starbursts. (The difference here
relates to the different mean bolometric luminosities between the two classes
of objects.) These predictions are an order of magnitude
brighter than what is seen in our objects (Table 4), with the possible exception of
the AGN IC 5063 where the minor axis filaments are very bright.
The higher predicted values may be due to a combination of several factors:
(i) the emitting regions have a small filling factor ($f < 1$), (ii) some dust is mixed in with 
the filaments (cf. Cecil\etal\ 2001), and (iii) $\xi < 1$ for the AGN/starburst fraction of the bolometric luminosity.

Thus we find that both AGNs and starbursts have the capability to
ionise much of the extended gas filaments observed far along the minor
axis, assuming that the radiation is not scattered or absorbed en
route. So we arrive at an important conclusion. {\it The distinction
between starburst and AGN outflow IDDs is one of timescale}, assuming
that the wind mechanism is not too dissimilar between the two sources
(e.g.\ Veilleux\etal\ 2005).  The time evolution of UV photon
production must be radically different in both cases.  To explore this
issue further, we examine more closely how the energy is deposited
into the surrounding medium.

\subsection{Time evolution of UV and mechanical luminosity in a starburst}

The starburst phase in a galaxy's life is a spectacular phenomenon
that can be observed to very high redshifts. Just how the formation of
hundreds of millions of stars can be triggered over the central
regions of galaxies in a dynamical time remains an open issue.  What
is well known is that the high mass stars evolve rapidly, giving off
prodigious amounts of UV radiation initially before entering into the
supernova phase, whereupon a huge amount of mechanical energy is
deposited into the ISM (Leitherer\etal\ 1992).

In Fig.~\ref{starburst99a}, we show the results from running {\it
Starburst99} with a conventional initial mass function (IMF)
for both instantaneous and continuous star formation models.
We show the time evolution of the ionizing and mechanical luminosities
arising from the starburst. Two upper mass cut-offs are presented in
order to illustrate the important contribution of the highest mass
stars. In Fig.~\ref{starburst99a}(a), the rapid drop in UV production
at 10 Myr is evident, as is the rise of mechanical output from
supernovae beyond here.  This corresponds to stars more massive than
13$M_\odot$ evolving off the main sequence to become supernovae.

Fig.~\ref{starburst99a}(a) can readily account for why stellar
photoionisation does not dominate over shock ionisation.
Thus, we conclude that the ionisation timescale ($T_I$) indicated
by starburst winds is such that $T_I \ga 10$ Myr.
Fig.~\ref{starburst99a}(b) effectively rules out the continuous star
formation model, at least for the starbursts in our sample. The model
predicts that stellar photoionisation is detectable at all times contrary to what
we observe.

In the next section, we show that the predicted mechanical luminosity
in the instantaneous star formation model is more than enough to
explain the observed shock ionisation.  Indeed, we argue that
Fig.~\ref{starburst99a} dates the observed starburst wind phenomenon
in the range 10 to 40 Myr.  A more precise age will depend on how the
starburst region evolved early in its life (Suchkov\etal\ 1994; Cooper
\etal\ 2008).  We discuss this issue in more detail below.

\subsection{Time evolution of UV and mechanical luminosity in an AGN}

We must now explain the dominance of AGN photoionisation of the
filaments in AGN wind objects. There are few reliable constraints on
the long-term temporal evolution of accretion-disk radiation
fields. Like starbursts, AGNs clearly display all the hallmarks of
duty cycles, bursts and outflows over the course of their lifetimes
(Soltan 1983; Cavaliere \& Padovani 1990). Most galaxies appear to
have central massive black holes, but only a few percent can be
classified as AGNs. The most likely explanation is that almost any
galaxy can display AGN activity if its nuclear regions experience
substantial gas accretion from outside.

But we arrive at the long-standing debate on what can be uniquely
attributed to the AGN as compared to, say, circumnuclear star-forming
activity (Terlevich\etal\ 1985). After all, the gas inflow that
triggers the former can just as easily give rise to the latter.  These
activities are commonly seen in close proximity, in particular, in
Seyfert galaxies. We address both of these issues now, i.e.\ why the
AGN dominates the photoionisation of the wind filaments, and whether
our conclusions are affected by the presence of the circumnuclear
starburst.

A reasonable AGN scenario is that the activity is initially triggered
by a gas dump, followed by a slow decline in the fuelling process
(e.g.  Norman \& Silk 1983) as the gas settles down within the nuclear
potential well. We assume an exponential model for the UV output from
an AGN such that
\begin{equation}
\label{time}
L_A/L_{\rm bol} = C_{\rm A, max} \exp(-t/t_A)
\end{equation}
where $C_{\rm A, max}$ is the maximum fraction of the bolometric luminosity
that the AGN achieves over the lifetime of the activity with half life $t_A$.
We further restrict this relation using the following normalization:
\begin{equation}
\label{normA}
\int_{\rm LL}^\infty {{L_A}\over{h\nu}}\ d\nu = {\cal N}_{\rm LyC,A}
\end{equation}
where LL denotes the Lyman limit ($h\nu=$ 13.6 eV) and the RHS is
given by equation~\ref{NLyCA}.  In practice, this normalization for
$C_{\rm A,max}$ and $t_A$ is difficult to do properly because our
observations constitute an instantaneous snapshot, rather than the
time-averaged behavior of its properties.  Of course, the same holds
true for starburst activity. In principle, IFU studies of a large
sample of AGNs and starbursts could provide accurate measurements of
$\xi_A$ and $\xi_S$ and therefore provide strong constraints on
$C_{\rm max}$ for both AGNs and starbursts.

If we integrate equation~\ref{time} over time ($\Delta t = 100$ Myr),
subject to the normalization in equation~\ref{normA}, we arrive at the
important result that $C_{\rm A,max} \propto t_A^{-1}$. The
normalization ensures that a range of AGN models depending on $C_{\rm
A,max}$ and $t_A$ produce the same ionizing radiation field averaged
over the lifetime of the source. In Fig.~\ref{starburst99b}, we
superimpose four such models on the instantaneous starburst model in
Fig.~\ref{starburst99a}.

As already mentioned, essentially all of our AGN sources show
appreciable circumnuclear star formation.  Since AGN photoionisation
is clearly dominant over stellar photoionisation in all of our AGN
sources, the predicted surface brightness values effectively rule out
AGN exponential timescales below 10 Myr. If the AGN UV radiation field
is ``beamed'' along the wind axis, in the sense that the AGN source is
{\it strongly} anisotropic, then shorter AGN timescales are
possible. But the dynamical timescale of the wind discussed in the
next section argues against this interpretation.  Furthermore, we have
been unable to find any reliable evidence for a UV boost along the
wind axis in AGNs.

But a question remains. In Fig.~\ref{starburst99b}, note how the
starburst mechanical luminosity can dominate over the AGN UV luminosity
beyond 10 Myr, assuming the mass/energy input can be efficiently
converted to ionizing radiation.  Why do we therefore not see the signatures
of shocks in AGN-driven winds?  Shocks are seen
in {\it starburst} winds because the central ionizing source fades rapidly
early in the evolution of the wind. But this problem only arises if we assume
that the mechanical energy in the AGN sources is supplied with a long delay
as for starburst-driven winds because, for example, the AGN wind is 
assisted by a circumnuclear starburst. The observations may therefore
indicate that the AGN-driven winds are fundamentally different from 
starburst-driven winds. In this scenario, the
mechanical energy is supplied by the accretion disk immediately, but
the shock ionizing luminosity never exceeds the luminosity of the 
central accretion disk.

We now consider the shock ionisation observed in starburst-driven winds.
We find that only a fraction of the available mechanical luminosity must be used in
shock-exciting the gas, either because the process is inefficient or because
the filling factor of the extended filaments is lower than unity.

\subsection{Dynamical timescale of a large-scale wind}

Here we address the dynamical timescale of galactic winds before going
onto derive   levels of optical emission powered by shock
ionisation.  Once the starburst or AGN fires up, it begins to deposit
energy into its surroundings.  This largely mechanical energy
over-pressurizes a circumnuclear cavity of hot gas that reaches a
temperature
\begin{equation}
T \approx 3 \times 10^8~\chi~\Lambda^{-1}~~{\rm K},
\label{eqn8}
\end{equation}
where $\chi$ is the efficiency of turning mechanical energy into heat,
and $\Lambda$ is the mass-loading factor (Veilleux\etal\ 2005). This
ratio measures the total mass of heated gas to the mass that is
directly ejected by SNe and stellar winds or by the AGN. Most galactic
winds appear to entrain gas from the interstellar medium as shown by
the rotation of this material about the flow axis (see
\citet{Veilleux_05} for a detailed discussion).  The entrained gas,
which can easily dominate the wind mass, is accounted for with
$\Lambda$ typically larger than unity (e.g.\ Suchkov\etal\ 1994).

The central cavity is too hot to cool efficiently and so builds up a pressure
\begin{equation}
P_o/k \sim 2 \times 10^6~\left({\dot{s}}\over{5\; M_\odot~{\rm yr}^{-1}}\right) \left( {R_*}\over{{\rm 1\; kpc}}\right)^{-2}~~{\rm 
K~cm}^{-3}
\label{eqn10}
\end{equation}
where $R_*$ is the radius of the energy injection zone.  This pressure
can significantly exceed the pressure of the undisturbed ISM. The hot
cavity then evolves like a stellar wind-blown bubble (e.g., Castor,
McCray, \& Weaver 1975; Koo \& McKee 1992a,b).  As hot gas expands
through the sonic radius, it cools adiabatically.  Beyond the sonic
radius, the wind drives a shock into the surrounding ISM and starts to
sweep up a shell of shocked gas, a process that gradually slows the
bubble to much less than the wind velocity. This marks the end of the
adiabatic ``free expansion'' phase, whose duration is set by the
mechanical luminosity of the starburst or AGN and the original density
of the ISM. After free expansion, the system develops distinct concentric zones
from the centre moving outwards due to the strong shocks (Cooper et al 2008; 
Dopita \& Sutherland 1996; Weaver\etal\ 1977).

\smallskip\noindent{\sl Energy-conserving bubbles.}  If radiative
losses of the overall system are negligible, the expanding bubble is
energy-conserving. In that case, the radius and the velocity of the
expanding shell of shocked ISM are given by (Castor\etal\ 1975;
Weaver\etal\ 1977)
\begin{equation}
r_{\rm shell} = 1.1~(\dot{E}_{44}/n_o)^{1/5}~t_6^{3/5}~~{\rm kpc},
\label{eqn12}
\end{equation}
and therefore
\begin{eqnarray}
V_{\rm shell} &=& 640~(\dot{E}_{44}/n_o)^{1/5}~t_6^{-2/5} \\
                        &=& 670~(\dot{E}_{44}/n_o)^{1/3}~r_{\rm shell, kpc}^{-2/3}~~{\rm 
km~s^{-1}}
\label{eqn13}
\end{eqnarray}
where $t_6$ is the age of the bubble in Myr, $n_o$ is the ambient
density in cm$^{-3}$. By convention, we use $\dot{E}_{44}$ for the mechanical 
luminosity
of the wind in units of 10$^{44}$ erg~s$^{-1}$ ($\dot{E}_{44}  \approx \xi L_{\rm bol}/
4\times 10^{11} L_\odot$).

\smallskip\noindent{\sl Momentum-conserving bubbles.}  If radiative
losses are significant, momentum conserving bubbles decelerate
somewhat faster such that $r_{\rm shell} \propto t^{\frac{1}{2}};
V_{\rm shell} \propto t^{-\frac{1}{2}} \propto r_{\rm shell}^{-1}$
(e.g., Avedisova 1972; Steigman, Strittmatter, \& Williams 1975; Koo
\& McKee 1992a,b).  For the galaxies we are considering here, the
inferred timescales are roughly comparable (see Table~\ref{bubble}).

\begin{table}[htdp]
\caption{A comparison of expansion time ($t_6$ in Myr) and shell velocity ($V_{\rm 
shell}$ in km s$^{-1}$) for a bubble
that has expanded to a radius $r_{\rm shell}$ (in kpc). Values are shown for an energy 
conserving and
a momentum conserving bubble.}
\begin{center}
\begin{tabular}{ccccc}
 & \multicolumn{2}{l}{energy conserving} & \multicolumn{2}{l}{momentum conserving}\\
              $r_{\rm shell}$           & $t_6$ & $V_{\rm shell}$ & $t_6$ & $V_{\rm shell}$ \\
                         
     &          &           &           &         \\
1  &  0.6  &  871   &  0.5   &  871 \\
2  & 1.7  &  549    & 2.0    & 616 \\
3  & 3.4  &  419   &  4.4   & 503 \\
4  & 5.6  &  346   &  7.8    & 435 \\
5  & 8.1  &  298   & 12.2    & 390 \\
\end{tabular}
\label{bubble}
\end{center}
\label{default}
\end{table}

In most simulations to date, the terminal velocity of gas entrained by
a galactic wind is $V_t \sim 400-800$ km s$^{-1}$ (e.g.\ Suchkov\etal\
1994; Cooper\etal\ 2008). Proper comparisons are difficult because the
models deal with clumpiness in the entrained gas, which the scaling
relations above do not.  Chevalier \& Clegg (1985) provide a useful
formula for the terminal velocity of a cloud accelerated by a wind
(see also Strickland \& Heckman 2008):
\begin{equation}
V_t \approx 430\; f_e^{1/4} f_m^{1/4} (r_o/{\rm 1\ kpc})^{-1/2} (\sigma_o/{10^{-3} {\rm g\ 
cm^{-2}}})^{-1/2}
\end{equation}
where $r_o$ and $\sigma_o$ are the initial distance and column density
of the cloud.  (This formula is specific to the ``constant velocity"
phase of the wind before it breaks free of the confining medium.)  The
factors $f_e$ and $f_m$ account for the fractions of supernova energy
and mass respectively that work to heat the diffuse gas. These authors
explore fractions in the range 0.1 to 1 and ultimately favour the high
end of this range. In any event, the dynamical time of the wind is
roughly $\tau_D \ge r_o/V_t \sim 2$ Myr (see also Cooper\etal\ 2008)
which is considerably longer than the cooling time of the entrained
gas.

The wind timescale is comparable to the dynamical time of the
starburst region itself, but is up to an order of magnitude less than
the ionisation timescale $T_I$ established by our analysis above.
This is a striking result that suggests a substantial delay ($\ga 10$
Myr) after the onset of star formation before the conditions for a
large-scale wind are properly established.

\subsection{Shock ionisation in a large-scale wind}

In the starburst wind objects, the overall ionisation properties are consistent with shock 
processes (see \S 3.1). A simple calculation can validate this. 
From Dopita \& Sutherland (1996), the total shock luminosity is
\begin{equation}
L_{\rm shock} \approx 2\times 10^{34} V_{\rm shock}^3 n_o A_{\rm shock} \; \; {\rm erg\ 
s^{-1}} .
\end{equation}
These authors find that about 50\% of the mechanical energy is converted into energy capable of ionizing gas.
Assuming there is sufficient gas to soak up the ionizing photons,
one can derive a useful relation for the H$\alpha$ luminosity, such that
\begin{equation}
L_{\rm shock} ({\rm H}\alpha) \approx 0.01 L_{\rm shock} V_{\rm shock}^{-0.6} \; \; {\rm 
erg\ s^{-1}} .
\end{equation}

We now derive a canonical H$\alpha$ surface brightness for shock-ionised filaments 
associated with the galactic wind. For a galaxy at a distance of $D= 10$ Mpc, the 
observed H$\alpha$ flux from the shock is 
\begin{eqnarray}
f_{\rm shock}({\rm H}\alpha) &\approx& 2.4\times 10^{-20} V_{\rm shock}^{2.4} n_o A_{\rm 
shock} \left({D}\over{{\rm 10\ Mpc}}\right)^{-2}\; \; \\
                                               & & \ \ \ \ \ \ \ \ \ \ \ \ \ \ \ \ \ \ {\rm erg\  cm^{-2}\ s^{-1}} 
\end{eqnarray}
which assumes, consistent with Dopita \& Sutherland (1996), that only a single shock surface
is observed with no correction for possible limb brightening.

The surface brightness of shocked emission at 1 kpc follows:
\begin{equation}
\label{surfbri2}
\mu_{\rm shock}({\rm H}\alpha) \approx 5.2\times 10^{-16} \left({V_{\rm shock}}\over{100\ 
\rm km\ s^{-1}}\right)^{2.4} \left({n_o}\over{\rm 10\ cm^{-3}}\right)  \; \; {\rm cgs}
\end{equation}
The pre-shock density normalization assumes that the halo gas in starburst galaxies follows an 
isothermal distribution (e.g.\ Cooper\etal\ 2008) normalised to a 
pre-shock density of $n_o = 10$ cm$^{-3}$ at 1 kpc consistent with material swept up in 
a wind (q.v. Chevalier \& Clegg 1985; see below). Post-shock densities inferred from the [SII] 
doublet (e.g.\ McCarthy\etal\ 1987) are typically in the low density limit
($< 100$ cm$^{-3}$) beyond $r=300$ pc. 

While equation~\ref{surfbri2} has a strong dependence on $V_{\rm
shock}$, the shock speed is well constrained by observation.  Some
authors have associated the terminal velocity of entrained gas with
the shock velocity induced by the wind (e.g.\ Maloney 1999). But the
levels of ionisation tell a different story. The strong [NII]/[OIII]
signal favours shock velocities of order 100 km s$^{-1}$ (e.g.\
Shopbell \& Bland-Hawthorn 1998; Veilleux \& Rupke 2002).  This is
supported by the soft x-ray emission which appears to be powered by
similar shock velocities. How are we to understand this?

In our picture, the fast hot wind accelerates denser gas to a terminal
velocity.  These clouds are ram-pressure confined by the fast wind. We
therefore write down a simple equation for shocks driven into the
clouds such that
\begin{equation}
n_w (V_w-V_t)^2 = n_o V_{\rm shock}^2
\label{Pram}
\end{equation}
where $n_w$ and $V_w$ are the wind density and velocity.

Strickland \& Heckman (2009) have recently conducted an excellent
re-evaluation of Chevalier \& Clegg's original model for the M82 wind.
The free-flowing or break-out wind speeds can be as high as $V_w =
3000$ km s$^{-1}$.  For our adopted values of $n_o$, $V_{\rm shock}$
and $V_t$ above, we can balance equation~\ref{Pram} with $n_w \sim
0.01$ cm$^{-3}$, a factor of a few higher than the canonical solution
at 1 kpc presented by Strickland \& Heckman (2009). We consider this
to be broadly consistent given the uncertainties. In our simple
analysis, the shock velocity is always several times less than the
terminal velocity of clouds caught in the flow.

Our simple model leads to an H$\alpha$ surface brightness of $\approx
5\times 10^{-16}$ cgs at 1 kpc. This approximate value is comparable
to what is presented in Table~\ref{gas extent} for the starburst
galaxies. In principle, the filling factor of the shocked material can
be low ($f<1$), and there will be moderate levels of extinction
internal to the entrained gas (i.e.\ $A_V < 0.3$ mag; Cecil\etal\
2001). But in principle, a moderately fast shock could account for the
observed, extended line emission.

We note that the H$\alpha$ surface brightness in
equation~\ref{surfbri2} is comparable to the observed AGN surface
brightness in Table~\ref{gas extent}, but much fainter than the
predicted AGN surface brightness in equation~\ref{surfbri1}.  This may
indicate that the filling factor in the wind filaments is $f\approx
0.1$.  Fortunately, a discussion of the {\it relative} merits of
different ionisation sources does not require knowledge of the gas
filling factor $f$. A lower filling factor affects the theoretical
prediction identically in equations~\ref{surfbri1} and \ref{surfbri2},
and usefully brings our models into line with the observations, at
least in principle.  Thus, if AGN winds involve similar processes to
starburst winds, we would not expect shocks to outshine direct
photoionisation from the AGN, consistent with our observations.

We have already argued that shocks outshine direct stellar
photoionisation from the starburst because the young starburst has
dimmed tremendously by the time the wind breaks out. Close to the
disk, we may still expect stellar photoionisation to dominate over
shocks owing to large-scale star formation in the disk (Bland-Hawthorn
\& Maloney 1999). But much beyond 300 pc or so, wind-driven shocks can
dominate the ionisation, as observed in our sample. If we invoke a
lower filling factor consistent with AGN winds, we can readily explain
NGC~3628, but now NGC~253 and NGC~1482 appear overluminous. Indeed, a
careful study of Table~\ref{gas extent} reveals the need to change $f$
over the range (0.1,1) to match our model to all of the
observations. This may indicate that our model is oversimplified and
needs to include a gas filling factor, or even a correction for
internal extinction (although this is found to be typically less than
10\% from the SPIRAL observations in the brightest regions).  We note
two things: (i) the inferred filling factors are not unreasonable and
can be directly measured in principle; (ii) alternatively, the
inferred extinction corrections are plausible, and these can also be
directly inferred from the Balmer decrement with data of sufficient
sensitivity.

\section{Sequence of events: the birth and evolution of galactic winds}
\label{events}

\smallskip
We conclude with a ``big picture'' model of the starburst wind phenomenon. Our purpose is to 
provide a physical explanation for the delay before the galactic wind gets going. We believe that
this simple model explains the broadest features of our new SPIRAL observations.
Therefore the stages of evolution for a galactic wind are as follows:

\smallskip
(1) Stars are born in dense molecular clouds whereupon the hot massive stars produce 
a strong UV radiation field (Lada \& Lada 2003);

\smallskip
(2) The mean free path of Lyman continuum photons is greatly extended by stellar 
winds (and radiation pressure) that blow bubbles into the surrounding ISM. (Good 
discussions of this process are to be found in Castor\etal\ 1975; Weaver\etal\ 1977; MacLow \& McCray 
1988; Tomisaka \& Ikeuchi 1988.) The intense UV radiation ionizes any pre-existing intracloud
medium but this may have been cleared out by an earlier wind.

\smallskip
(3) The over-pressured warm gas evaporated from the surfaces of dense clouds
helps to fill the wind-blown ``voids'' and intracloud region. The connection between 
the evaporating cloud skins and the intracloud wind material may help the wind to entrain
denser gas from these same clouds (Cooper\etal\ 2008).

\smallskip
(4) The bulk of the supernova ejecta arises after 8-10 Myr
(Fig.~\ref{starburst99a}; Leitherer\etal\ 1992). As noted by others (e.g.\ Chevalier \&
Clegg 1985), the ejecta ultimately thermalizes the diffuse material which
escapes as a wind; any shock heating of the denser medium escapes
as infrared radiation (Strickland \& Stevens 2000). The UV intensity declines
dramatically since many of the high-mass stars have now exploded.

\smallskip
Thus we are led to an important conclusion. {\it The large-scale wind
cannot become established until vast quantities of warm diffuse gas are
dislodged from the surfaces of dense clouds by UV evaporation.}  The
ionised material with a sufficiently long cooling time is then
mechanically heated by the supernova mass/energy input. It is
therefore not surprising that, by the time the wind gets going, most
of the short-lived massive stars have faded or disappeared. We would
therefore {\it not} expect to see a strong signature of stellar
photoionisation in the extended wind filaments if the starburst 
phenomenon arises from a {\it single} burst of star formation. 
{\it Our new observations may provide the strongest evidence yet that
starbursts truly are impulsive events}. The fact that the
signature of AGN photoionisation {\it is} clearly visible in the wind
filaments argues that the accretion-disk UV timescale is considerably
longer than the starburst timescale. Furthermore, unlike the starburst,
the mechanical energy input {\it must} have occurred early in the life of 
the AGN, presumably at a time when jets can also be launched.
We argue that our observations
are a testament to this simple picture.

Conversely, if there are starburst sources where stellar
photoionisation {\it does} dominate at large radial distances (cf. Shopbell
\& Bland-Hawthorn 1998), we anticipate that the circumnuclear regions
in these sources are characterised by an extended episode of star
formation (i.e.\ continuous star formation) due to an extended period
of fuelling and accretion (e.g.\ sustained merger activity).

\section{Summary}
\label{discussion}

\smallskip
The role of mass/energy feedback and recycling in galactic winds over
cosmic time is very poorly understood. In particular, we are far from
a detailed understanding of how gas accretion to the circumnuclear
regions fuels local star formation and nuclear activity. So what can
we say?

The galaxy wind phenomenon is well established and its influence can
be detected in the highest redshift sources (Heckman\etal\ 1990;
Strickland\etal\ 2004b; Veilleux\etal\ 2005). Galactic winds are
associated with both star formation (e.g.\ starbursts), nuclear
activity (e.g.\ FR~I radio sources), and a combination of star
formation and nuclear activity (e.g.\ Seyfert galaxies). But this
raises many questions.  How do winds get going, how energetic are
they, how long do they last, and what is their role in galaxy
evolution? Do we arrive at similar answers for AGN-driven winds and
starburst-driven winds?

Here we are able to shed light on the first question (\S
\ref{events}). We have carried out the most detailed and extensive
survey to date of extended plasma gas in active galaxies. We
constructed ionisation diagnostic diagrams to study the nature of the
emission in ten galaxies. We find that the wholly starburst-powered
sources have extended emission that is largely consistent with shock
ionisation. This sets an important lower limit on the time at which
the starburst phenomenon began ($\le 50$ Myr). At this time, the
stellar UV ionisation sources have long since faded from view which
essentially rules out a continuous star formation model for the
nucleus, at least in our limited subset of starburst galaxies. An
interesting question is whether we can identify those sources that are
in an early starburst phase and before the onset of the large-scale
wind.

The second question presents difficulties for us, at least in the
present analysis.  In all of our AGN sources, there is evidence of
circumnuclear star formation: the present data are unable to establish
which source of activity is responsible for the wind. In future
studies, it may be possible to shed light on this question in one of
two ways. X-ray spectroscopy can, in principle, establish the
[$\alpha$/Fe] abundance ratio of the outflowing gas. Enhanced $\alpha$
abundances implicate the presence of SN II, and therefore the central
importance of the starburst in driving the outflow.  Such an effect
may have already been seen in dwarf starburst galaxies
(\citet{Martin02}, Ott, Walter \& Brinks (2005)), but we are unaware
of similar studies in Seyfert outflow sources. This is an important
avenue for future research.

An alternative approach is to carefully trace the pressure changes
over the circumnuclear regions down to small galactic radii.
O'Connell \& Mangano (1978) noticed that the gas densities, as
measured from the [SII] doublet, appear to decrease along the minor
axis of M82. Chevalier \& Clegg (1985) interpreted this observation as
a change of wind pressure in a starburst-driven wind. This idea has
been explored extensively by Heckman\etal\ (1990) who used the \Sii\
doublet to measure the electron densities as a function of radius in a
subset of starburst galaxies. In M82, high pressures have been
inferred from the \Siii\ doublet (Houck\etal\ (1984),
\citet{Smith06}), from fine structure lines in the far infrared
(Lugten\etal\ 1986; Duffy\etal\ 1987), and from the non-thermal radio
continuum (Schaaf\etal\ 1989). After certain assumptions, pressures
can be inferred by combining observations of radio recombination lines
and the associated free-free emission (Rodriguez \& Chaisson 1980;
Seaquist, Bell \& Bignell 1985). The intrinsically higher resolution
of linked radio receivers should allow the pressure profile to be
traced to the smallest VLBI spatial scales. Naively, the pressure
profile can be relatively steep closer to the black hole compared to
an anticipated flatter profile across the energy injection region of a
starburst (Chevalier \& Clegg 1985).  Thus, we may be able to
distinguish between AGN-powered vs. starburst-powered momentum by
measuring the gas pressure as a function of galactocentric radius.  We
leave this to future work.

In the case of the AGN sources, we find that the extended line
emission is always powered by the accretion disk rather than by shock
ionisation. Some of these sources are clearly associated with central
starbursts that may well contribute momentum to the outflow in which
case age-dating the onset of the outflow is more problematic. In the
case of the AGN-driven outflow, these could have commenced as recently
as 10 Myr or so, although if the starburst dominates, the longer
timescale is preferred. The latter case is interesting because it
would indicate that the AGN source has a lifetime approaching 100 Myr,
and that the amount of gas converted to energy is of order
10$^8\;M_\odot$.

In the next paper, we combine the ionisation diagnostics with the
emission-line kinematics to learn more about the energetics of the
outflowing gas. The integral field spectrograph approach is clearly
very powerful for studying extended emission line sources.  We stress
that our methods for deriving and comparing ionisation and dynamic
timescales can be extended to a wider class of active galaxies (e.g.\
quasar hosts), mergers (e.g.\ ULIRGs) and star-forming galaxies (e.g.\
dwarf irregulars). It will be interesting to compare the inferred
timescales with the star formation histories extracted from
circumnuclear stellar populations. In high-redshift galaxies where
stellar populations are harder to observe, the ionisation diagnostics
may provide key information.  In the Appendix, we include a brief
discussion about how future studies can be improved to advance this
work further.

\smallskip
\acknowledgements {We are indebted to an anonymous referee for comments
which improved the clarity of our text. We gratefully
acknowledge the superb efforts of the operational and support
staff of the AAT staff, without who these observations would not have
been possible.  We thank Ryan Cooke for his help during some of the
observations.  JBH is supported by a Federation Fellowship from the
Australian Research Council.  This research has made use of the
NASA/IPAC Extragalactic Database (NED) which is operated by the Jet
Propulsion Laboratory, California Institute of Technology, under
contract with the National Aeronautics and Space Administration.}

\smallskip
{\it Facilities:} \facility{AAT (AAOmega-SPIRAL)}

\clearpage

\clearpage

\clearpage

\clearpage

\clearpage

\begin{figure*}
\begin{center}
\includegraphics[width=55mm]{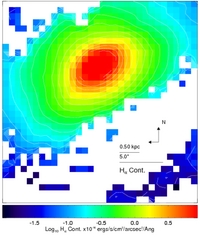}	    
\includegraphics[width=55mm]{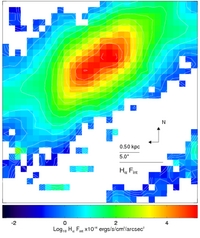}
\includegraphics[width=55mm]{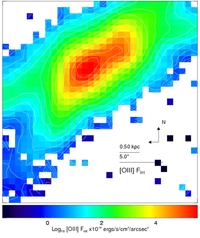}\\
\includegraphics[width=55mm]{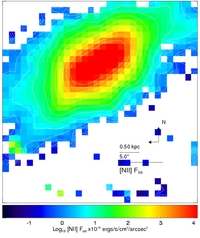}		 
\includegraphics[width=55mm]{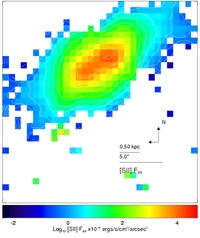}		 
\end{center}
  \caption{\label{ic5063 lines} IC~5063: Emission line intensity maps
  following Fig.~\ref{ngc1482 lines}.}
\end{figure*}

\begin{figure*}
\begin{center}
\includegraphics[width=55mm]{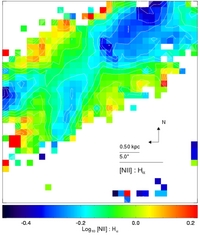}
\includegraphics[width=55mm]{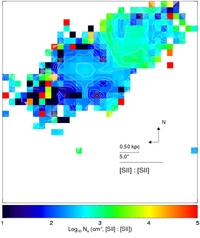}\\
\includegraphics[width=55mm]{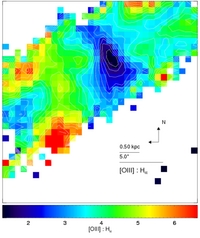}
\includegraphics[width=55mm]{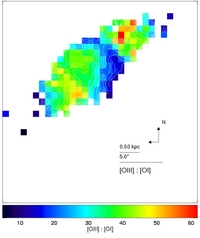}
\end{center}
  \caption{\label{ic5063 ratios} IC5063: Line ratio maps are
  constructed following Fig.~\ref{ngc1482 ratios}.}
\end{figure*}

\begin{figure*}
\begin{center}
\includegraphics[width=50mm]{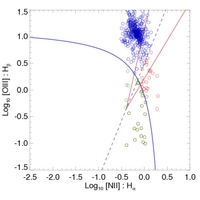}
\includegraphics[width=50mm]{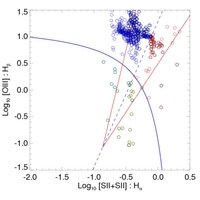}
\includegraphics[width=50mm]{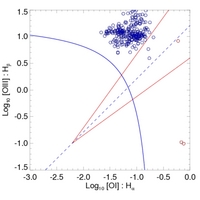}\\
\includegraphics[width=50mm]{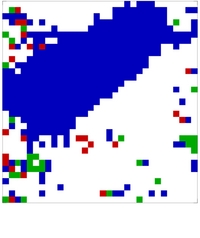}
\includegraphics[width=50mm]{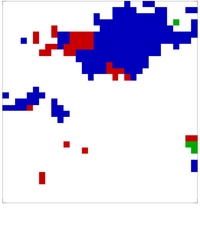}
\includegraphics[width=50mm]{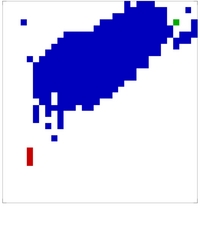}
\end{center}
  \caption{\label{ic5063 VO} IC5063: Ionisation Diagnostic Diagrams
  are constructed following Fig.~\ref{ngc1482 VO}.}
\end{figure*}

\clearpage

\begin{figure}
\includegraphics[width=120mm]{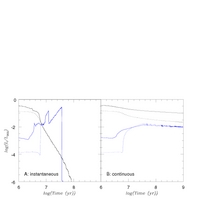}		 
\caption{\label{starburst99a} Ionizing luminosity (black)
vs.\ mechanical luminosity (blue) evolution in two different star
forming scenarios: (A) instantaneous burst; (B) continuous. In both
cases, a Salpeter initial mass function is adopted but with two
different upper mass cut-offs: 30 M$_\odot$ (dotted line), 100
M$_\odot$ (solid line).  }
\end{figure}

\begin{figure}
\includegraphics[width=120mm]{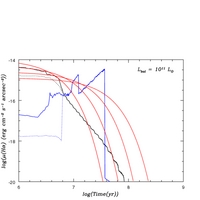}
\caption{\label{starburst99b}Predicted evolution of H$\alpha$ surface
brightness of optically thick gas at 1 kpc from a central active
source with $L_{\rm bol} = 10^{11} L_\odot$. The black and blue curves
are described in Fig.~\ref{starburst99a}. The red curves correspond to
4 different sets of $(C_{\rm A,max};t_A)$ in our AGN model, i.e.\ (0.1,
20), (0.2, 10), (0.4, 5), (0.8, 2.5) where the second value in each
pair is the exponential timescale in Myr; larger values therefore
produce slower decays. The predicted values are at least an order of
magnitude higher than what we observe as we discuss in \S 4. We do
not include the evolution of the AGN's mechanical luminosity for comparison
because this is highly uncertain. But since AGN photoionization seems to
dominate over shocks, we must assume that the mechanical 
luminosity profile falls below the AGN UV curve at all times and therefore
does not resemble the starburst profile.}
\end{figure}

\appendix

\section{The use of integral field spectrographs -- lessons learned}

The use of a double-beam integral field spectrograph has been critical
to the success of this project. It provides both detailed ionisation
diagnostics and kinematics over the extent of the outflowing gas. We
have demonstrated how these IDDs can be used to classify the nature of
the wind outflow for each source, and to define spatial zones in which
different ionisation sources dominate.  Furthermore, we have begun the
process of spectral and spatial kinematic and ionisation deconvolution
for each source, in an effort to disentangle the complexity of each
system.

We conclude that blue sensitivity (below 500nm) is a key factor to
many of the diagnostic requirements, in particular sensitivity to the
\Hb\ and \Oiii\ spectral region (with SPIRAL we can not comment on the
utility of \Oii\ emission feature for such work).  As a clear shock
diagnostic, \Oi\ is also of great value to classification of the
nature of the ionisation source, although in all cases presented in
this work, sensitivity to the \Oiii\ emission still presents the
primary limitation.  This limitation is imposed in part by the reduced
throughput of the blue arm of the AAOmega system \citep{Sharp06} but
will remain an important consideration for more balanced systems.

Spectral resolution is critical to disentangling the multiple emission
components of many of the sources discussed, with line splitting in
many of the outflows only marginally resolved in our R$\sim$5000
spectroscopy ($\sim$50\kms).  One must of course also consider
wavelength coverage.  While many of the fainter lines common to ISM
studies may be beyond the reach of a survey such as ours (for example
the temperature sensitive \Nii\ $\lambda$5756 line is typically rather
faint) we have demonstrated the added value of an extended spectral
range in contrast to the limited coverage usually practical with
Fabry-P\'erot tunable filter systems such as that employed for example
by \citet{ttf03}.  A range of source redshifts, even for a nearby
galaxy sample, further complicates the choice of wavelength range
selection.

The observations and interpretation presented here have been possible
primarily due to the high surface brightness sensitivity of the
AAOmega-SPIRAL system afforded by the large projected spatial pixel
size on the sky.  The 0.7arcsec on a side square elements provide an
individual field of view of 0.49arcsec$^2$, which represents factors
of $\sim$2, 5.4 \& 12 increase in contrast to 0.5,0.3 or 0.2arcsec
systems, more than accounting for the smaller size of the prime mirror
of the AAT 3.9m when compared to systems available on the current
generation of large optical telescopes.  This allows SPIRAL to move
out of the detector readnoise limit, below which re-binning of
adjacent apertures will not recover signal-to-noise quickly.  As
recently demonstrated by Westmoquette \etal\ 09a/b, in truth one must
ultimately combine the wide field of view and low surface brightness
observations practicable with large pixels, with dedicated narrow
field observations at higher spatial resolution of critical regions of
each object.

\end{document}